\documentstyle[preprint,aps,floats,psfig]{revtex}
\catcode`@=11
\def\references{%
\ifpreprintsty
\bigskip\bigskip
\hbox to\hsize{\hss\large \refname\hss}%
\else
\vskip24pt
\hrule width\hsize\relax
\vskip 1.6cm
\fi
\list{\@biblabel{\arabic{enumiv}}}%
{\labelwidth\WidestRefLabelThusFar  \labelsep4pt %
\leftmargin\labelwidth %
\advance\leftmargin\labelsep %
\ifdim\baselinestretch pt>1 pt %
\parsep  4pt\relax %
\else %
\parsep  0pt\relax %
\fi
\itemsep\parsep %
\usecounter{enumiv}%
\let\p@enumiv\@empty
\def\theenumiv{\arabic{enumiv}}%
}%
\let\newblock\relax %
\sloppy\clubpenalty4000\widowpenalty4000
\sfcode`\.=1000\relax
\ifpreprintsty\else\small\fi
}
\catcode`@=12

\def\lsim{\mathrel{\raise.3ex\hbox{$<$\kern-.75em\lower1ex\hbox{$\sim$}}}}
\def\gsim{\mathrel{\raise.3ex\hbox{$>$\kern-.75em\lower1ex\hbox{$\sim$}}}}
\begin{document}
\tightenlines

%\font\fortssbx=cmssbx10 scaled \magstep2
%\hbox to \hsize{{\fortssbx University of Wisconsin - Madison}
\hfill\vtop{
\hbox{AMES-HET-01-12}
\hbox{BUHEP-01-32}
\hbox{MADPH-01-1250}
\hbox{hep-ph/0112119}
\hbox{}}

\vspace*{.25in}
\begin{center}
{\large\bf Breaking Eight--fold Degeneracies in Neutrino \\
$CP$ Violation, Mixing, and Mass Hierarchy}\\[10mm]
V. Barger$^1$, D. Marfatia$^2$ and K. Whisnant$^3$\\[5mm]
\it
$^1$Department of Physics, University of Wisconsin,
Madison, WI 53706, USA\\
$^2$Department of Physics, Boston University,
Boston, MA 02215, USA\\
$^3$Department of Physics and Astronomy, Iowa State University,
Ames, IA 50011, USA

\end{center}
\thispagestyle{empty}

\begin{abstract}

\vspace*{-.35in}

\noindent

We identify three independent two-fold parameter degeneracies
$(\delta\,, \theta_{13})$, sgn$(\delta m^2_{31})$ and $(\theta_{23}\,,
\pi/2-\theta_{23})$ inherent in the usual three--neutrino analysis of
long--baseline neutrino experiments, which can lead to as much as an
eight--fold degeneracy in the determination of the oscillation
parameters. We discuss the implications these degeneracies have for
detecting $CP$ violation and present criteria for breaking them. A
superbeam facility with a baseline at least as long as the distance
between Fermilab and Homestake (1290 km) and a narrow band beam with
energy tuned so that the measurements are performed at the first
oscillation peak can resolve all the ambiguities other than the
$(\theta_{23}\,, \pi/2-\theta_{23})$ ambiguity (which can be resolved at
a neutrino factory) and a residual $(\delta\,, \pi-\delta)$
ambiguity. However, whether or not $CP$ violation occurs in the neutrino
sector can be ascertained independently of the latter two
ambiguities. The $(\delta\,,\pi-\delta)$ ambiguity can be eliminated by
performing a second measurement to which only the $\cos \delta$ terms
contribute. The hierarchy of mass eigenstates can be determined at other
oscillation peaks only in the most optimistic conditions, making it
necessary to use the first oscillation maximum. We show that the
degeneracies may severely compromise the ability of the proposed
SuperJHF--HyperKamiokande experiment to establish $CP$ violation.  In
our calculations we use 
approximate analytic expressions for oscillation probabilitites
that agree with numerical solutions with a realistic Earth density profile.

%; we show
%that for $E_\nu > 0.5$~GeV and distances of 4000~km or less they agree
%well with numerical solutions with a realistic Earth density profile.

\end{abstract}

\newpage

\section{Introduction}

The up/down asymmetry of the neutrino flux (originating from cosmic ray
interactions with the atmosphere) at SuperKamiokande is now
a 10$\sigma$ effect. A compelling interpretation of this result is that
neutrinos have mass and oscillate from one flavor to another. The
atmospheric neutrino deficit is explained as a consequence of $\nu_\mu
\rightarrow \nu_\tau$ oscillations with almost maximal amplitude and
mass-squared difference, $\delta m^2_{31} \sim 3 \times 10^{-3}$
eV$^2$~\cite{Toshito:2001dk}. The K2K experiment~\cite{Hill:2001gu} with
a baseline of 250 km has preliminary results that are in agreement with
this interpretation. Oscillations of $\nu_\mu$ to $\nu_e$ as an
explanation of the atmospheric anomaly are ruled out by the
CHOOZ~\cite{CHOOZ} and Palo Verde~\cite{paloverde} reactor experiments,
which place a bound on the amplitude smaller than 0.1 at the 95\%
C.L. in the $\delta m^2_{31}$ region of interest. The
MINOS~\cite{minos}, ICARUS~\cite{icarus} and OPERA~\cite{opera}
experiments are expected to come online in 2005 and study aspects of the
oscillations at the atmospheric scale~\cite{Barger:2001yx}. The low
energy beam at MINOS will allow a very accurate determination of the
leading oscillation parameters. ICARUS and OPERA should provide concrete
evidence that $\nu_\mu \rightarrow \nu_\tau$ oscillations are
responsible for the atmospheric neutrino deficit by identifying tau
neutrino events.

Measurements of electron neutrinos from the Sun also provide strong
evidence for neutrino oscillations. The flux of electron neutrinos from
the Sun observed in several different experiments is smaller than the
Standard Solar Model~\cite{Bahcall:2000nu} (SSM) prediction by a factor
of 1/3--1/2.  
The recent SNO charged-current
measurements show that $\nu_e \rightarrow \nu_{\mu\,,\tau}$ oscillations
explain the $\nu_e$ flux suppression relative to the
SSM~\cite{Ahmad:2001an}. The solution with a large mixing angle (LMA)
and small matter effects ($\delta m^2_{21} \sim 5 \times 10^{-5}$ eV$^2$
and amplitude close to 0.8) has emerged as the most likely solution to
the solar neutrino problem~\cite{lma}.  This solution will be tested
decisively by the KamLAND reactor neutrino experiment~\cite{kamland}.

There are several parameter
degeneracies that enter the determination of the neutrino mixing matrix
which can be removed only with future oscillation studies with
superbeams or neutrino factories. See Table~\ref{tab:baselines} for a
sample of proposed baselines.
 \begin{table}[b]
\squeezetable
\caption[]{Baseline distances in km for some detector sites (shown in
parentheses) for neutrino beams from Fermilab, Brookhaven, JHF, and CERN.}
\label{tab:baselines}
\begin{tabular}{llll}
\multicolumn{4}{c}{Beam source}\\
Fermilab & Brookhaven & JHF & CERN\\
\hline
\hline
&&& \phantom{1}150 (Frejus) \\
\hline
& \phantom{1}350 (Cornell) & \phantom{1}295 (Super--K) & \\
\hline
\phantom{1}730 (Soudan) & & & \phantom{1}730 (Gran Sasso)\\
\hline
1290 (Homestake) & & 1200 (Seoul) & \\
\hline
1770 (Carlsbad) & 1720 (Soudan) & & \\
\hline
& & 2100 (Beijing) & \\
\hline
2640 (San Jacinto) & 2540 (Homestake) & & \\
\hline
2900 (SLAC) & 2920 (Carlsbad) & & \\
\end{tabular}
\end{table}
A notable
example is the $U_{e3}$ (=$\sin \theta_{13}$ e$^{-i\delta}$) element. Only
an upper bound exists on $\theta_{13}$, nothing is presently known about the
$CP$ phase $\delta$, and the two always appear in combination in the mixing
matrix. It is the breaking of such degeneracies that will be of concern to 
us in this work. 

In Section II we identify all the potential parameter degeneracies in
the mixing matrix.  We restrict our attention to the $3\times 3$ matrix
that describes the mixing of active neutrinos, setting aside the
possibility that the atmospheric, solar and
LSND~\cite{Athanassopoulos:1997pv} data may require the existence of a
fourth neutrino that is sterile.  The parameter ambiguities are
connected with not only neutrino mixing but also the neutrino mass
pattern; we pay particular attention to the implication of these
ambiguities for the detection of $CP$ violation. In Section III, within
the context of a superbeam experiment~\cite{sup}, we present methods by
which all but one of these degeneracies can be resolved, and argue that
the remaining ambiguity can be settled at a neutrino
factory~\cite{geer}. We also discuss the implications of the
degeneracies on the proposed SuperJHF--HyperKamiokande
experiment~\cite{jhfsk}, which would have a 4~MW proton driver and a
1~Mt water cerenkov detector (40 times larger than SuperKamiokande).
We summarize our results in Section IV. In an
appendix we provide a complete set of approximate analytic expressions
for the oscillation probabilities that are useful for superbeams and
neutrino factories, and define their domain of validity by making
comparisons with numerical solutions of the evolution equations.

\section{Parameter degeneracies}
\label{sec:degeneracies}

In this section we identify the three types of parameter
degeneracies that can occur in the three--neutrino framework when
$\nu_\mu \to \nu_e$ and $\bar\nu_\mu \to \bar\nu_e$ oscillation probabilities  
are used to extract
the neutrino parameters. We use approximate
formulas~\cite{cervera,freund} for neutrino propagation in matter of
constant density to illustrate the degeneracies. In each case we
discuss the implications for detecting $CP$ violation.

\subsection{Oscillation probabilities in matter}
\label{sec:prob}

The neutrino flavor eigenstates $\nu_\alpha\ (\alpha = e, \mu, \tau)$
are related to the mass eigenstates $\nu_j\ (j = 1, 2, 3)$ in vacuum by
\begin{equation}
\nu_\alpha = \sum_j U^{*}_{\alpha j} \nu_j \;,
\end{equation}
where $U$ is a unitary $3\times3$ mixing matrix. The propagation of
neutrinos through matter is described by the evolution
equation~\cite{matter,kuo}
\begin{equation}
i{d\nu_\alpha\over dx} = \sum_\beta \left( \sum_j U_{\alpha j}
U^*_{\beta j}
{m^2_j\over 2E_\nu} + {A\over 2E_\nu} \delta_{\alpha e} \delta_{\beta e}
\right) \nu_\beta \;,
\label{eq:evolution}
\end{equation}
where $x = ct$ and $A/2E_\nu$ is the amplitude for coherent forward
charged-current $\nu_e$ scattering on electrons,
\begin{equation}
A = 2\sqrt 2\, G_F \, N_e \, E_\nu = 1.52\times10^{-4}\,{\rm eV^2}
Y_e \, \rho\,({\rm g/cm^3}) E_\nu\,(\rm GeV) \;,
\label{eq:A}
\end{equation}
Here $N_e$ is the electron number density, which is the product of
the electron fraction $Y_e(x)$ and matter density $\rho(x)$.
In the Earth's crust and mantle the average matter density is typically
3--5~g/cm$^3$ and $Y_e\simeq 0.5$. The propagation equations can be
re-expressed in terms of mass-squared differences
\begin{equation}
i{d\nu_\alpha\over dx} = \sum_\beta {1\over 2E_\nu} \left( \delta m_{31}^2
U_{\alpha3} U_{\beta3}^* + \delta m_{21}^2 U_{\alpha2} U_{\beta2}^* + A
\delta_{\alpha e} \delta_{\beta e}\right) \;,
\end{equation}
where $\delta m_{jk}^2 \equiv m_j^2 - m_k^2$. The neutrino mixing matrix
$U$ can be specified by 3 mixing angles ($\theta_{23}, \theta_{12},
\theta_{13}$) and a $CP$-violating phase $\delta$. We adopt the
parameterization
\begin{equation}
U
= \left( \begin{array}{ccc}
  c_{13} c_{12}       & c_{13} s_{12}  & s_{13} e^{-i\delta} \\
- c_{23} s_{12} - s_{13} s_{23} c_{12} e^{i\delta}
& c_{23} c_{12} - s_{13} s_{23} s_{12} e^{i\delta}
& c_{13} s_{23} \\
    s_{23} s_{12} - s_{13} c_{23} c_{12} e^{i\delta}
& - s_{23} c_{12} - s_{13} c_{23} s_{12} e^{i\delta}
& c_{13} c_{23} \\
\end{array} \right) \,,
\label{eq:MNS}
\end{equation}
where $c_{jk} \equiv \cos\theta_{jk}$ and $s_{jk} \equiv
\sin\theta_{jk}$. In the most general $U$, the $\theta_{ij}$ are
restricted to the first quadrant, $0\le \theta_{ij} \le \pi/2$,
with $\delta$ in the range $0 \le \delta < 2\pi$. We assume that
$\nu_3$ is the neutrino eigenstate that is separated from the other two,
and that the sign of $\delta m^2_{31}$ can be either positive or
negative, corresponding to the case where $\nu_3$ is either above or
below, respectively, the other two mass eigenstates. The magnitude of
$\delta m^2_{31}$ determines the oscillation length of atmospheric
neutrinos, while the magnitude of $\delta m^2_{21}$ determines the
oscillation length of solar neutrinos, and thus $|\delta m^2_{21}| \ll
|\delta m^2_{31}|$. 
%If we allow the sign of $\delta m^2_{21}$ to be
%positive or negative 
%and we can restrict $\theta_{12}$ to the range $[0,\pi/4]$. 
If we accept the likely conclusion that the solar solution is
LMA~\cite{lma}, then $\delta m^2_{21} > 0$ and we can restrict
$\theta_{12}$ to the range $[0,\pi/4]$.  It is known from reactor
neutrino data that
$\theta_{13}$ is small, with $\sin^22\theta_{13} \le 0.1$ at the  95\%
C.L.~\cite{CHOOZ}. Thus a set of parameters that unambiguously spans the
space is $\delta m^2_{31}$ (magnitude and sign), $\delta m^2_{21}$, 
$\sin^22\theta_{12}$, $\sin\theta_{23}$, and $\sin^22\theta_{13}$; only
the $\theta_{23}$ angle can be below or above $\pi/4$.

In the context of three-neutrino models the usual method proposed for detecting
$CP$ violation in long-baseline experiments with a conventional neutrino
beam is to measure the oscillation channels $\nu_\mu \to \nu_e$ and
$\bar\nu_\mu \to \bar\nu_e$ (or $\nu_e \to \nu_\mu$ and $\bar\nu_e \to
\bar\nu_\mu$ for a neutrino factory). Both leading and subleading
oscillation contributions must be involved and the oscillations must be
non--averaging for $CP$--violation effects~\cite{1980CP}. For
illustrative purposes we use the constant density matter approximation,
although in an exact study variations of the density along the neutrino
path should be implemented. Approximate formulas for the oscillation
probabilities in matter of constant density in the limit $|\delta
m^2_{21}| \ll A, |\delta m^2_{31}|$ already exist in the
literature~\cite{cervera,freund}. We adopt the form in
Ref.~\cite{freund}, where $\theta_{13}$ is also treated as a small
parameter and the mixing angles in matter are found in terms of an
expansion in the small parameters $\theta_{13}$ and $\delta m^2_{21}$.
We introduce the notation
\begin{eqnarray}
\Delta &\equiv& |\delta m_{31}^2| L/4E_\nu
= 1.27 |\delta m_{31}^2/{\rm eV^2}| (L/{\rm km})/ (E_\nu/{\rm GeV}) \,,
\label{eq:D}\\
\hat A &\equiv& |A/\delta m_{31}^2| \,,
\label{eq:Ahat}\\
\alpha &\equiv& |\delta m^2_{21}/\delta m^2_{31}| \,.
\label{eq:alpha}
\end{eqnarray}
Up to second order in $\alpha$ and $\theta_{13}$, the oscillation
probabilities for $\delta m^2_{31} > 0$ and $\delta m^2_{21} > 0$ are
\begin{eqnarray}
P(\nu_\mu \to \nu_e) =
x^2 f^2 + 2 x y f g (\cos\delta\cos\Delta - \sin\delta\sin\Delta)
+ y^2 g^2\,,
\label{eq:P}\\
\bar P(\bar\nu_\mu \to \bar\nu_e) =
x^2 \bar f^2 + 2 x y \bar f g (\cos\delta\cos\Delta
+ \sin\delta\sin\Delta) + y^2 g^2 \,,
\label{eq:Pbar}
\end{eqnarray}
respectively, where
\begin{eqnarray}
x &\equiv& \sin\theta_{23} \sin 2\theta_{13} \,,
\label{eq:x}\\
y &\equiv& \alpha \cos\theta_{23} \sin 2\theta_{12} \,,
\label{eq:y}\\
f, \bar f &\equiv& \sin((1\mp\hat A)\Delta)/(1\mp\hat A) \,,
\label{eq:f}\\
g &\equiv& \sin(\hat A\Delta)/\hat A \,.
\label{eq:g}
\end{eqnarray}
The coefficients $f$ and $\bar f$ differ due to matter effects ($\hat A
\ne 0$). To obtain the probabilites for $\delta m^2_{31} < 0$, the
transformations $\hat A \to - \hat A$, $y \to -y$ and $\Delta \to
-\Delta$ (implying $f \leftrightarrow -\bar f$ and $g \to -g$) can be
applied to the probabilities in Eqs.~(\ref{eq:P}) and (\ref{eq:Pbar}) to
give
\begin{eqnarray}
P(\nu_\mu \to \nu_e) =
x^2 \bar f^2 - 2 x y \bar f g (\cos\delta\cos\Delta
+ \sin\delta\sin\Delta) + y^2 g^2\,,
\label{eq:P2}\\
\bar P(\bar\nu_\mu \to \bar\nu_e) =
x^2 f^2 - 2 x y f g (\cos\delta\cos\Delta
- \sin\delta\sin\Delta) + y^2 g^2 \,.
\label{eq:Pbar2}
\end{eqnarray}
For a $T$-reversed channel, the corresponding probabilities are found by
changing the sign of the $\sin\delta$ term. In Eqs.~(\ref{eq:P}),
(\ref{eq:Pbar}), (\ref{eq:P2}), and (\ref{eq:Pbar2}) we have assumed
$\delta m^2_{21} > 0$, which is what one expects for the LMA solar
solution; for $\delta m^2_{21} < 0$, the corresponding formulae are
obtained by $y \to -y$. These expressions are accurate as long as $\theta_{13}$
is not too large, and they are valid at $E_\nu > 0.5$ GeV ($\hat A \gsim
0.04 (3\times10^{-3}$~eV$^2/|\delta m^2_{31}|)$) for all values of
$\delta m^2_{21}$ currently favored by solar neutrino
experiments. We
expand on the domain of validity of these equations 
in the Appendix. The
corresponding expansion in $\alpha$ and $\theta_{13}$ in a vacuum can
be found by the substitutions
$f, \bar f, g\to \sin\Delta$.

\begin{figure}[h!]
\centering\leavevmode
\psfig{file=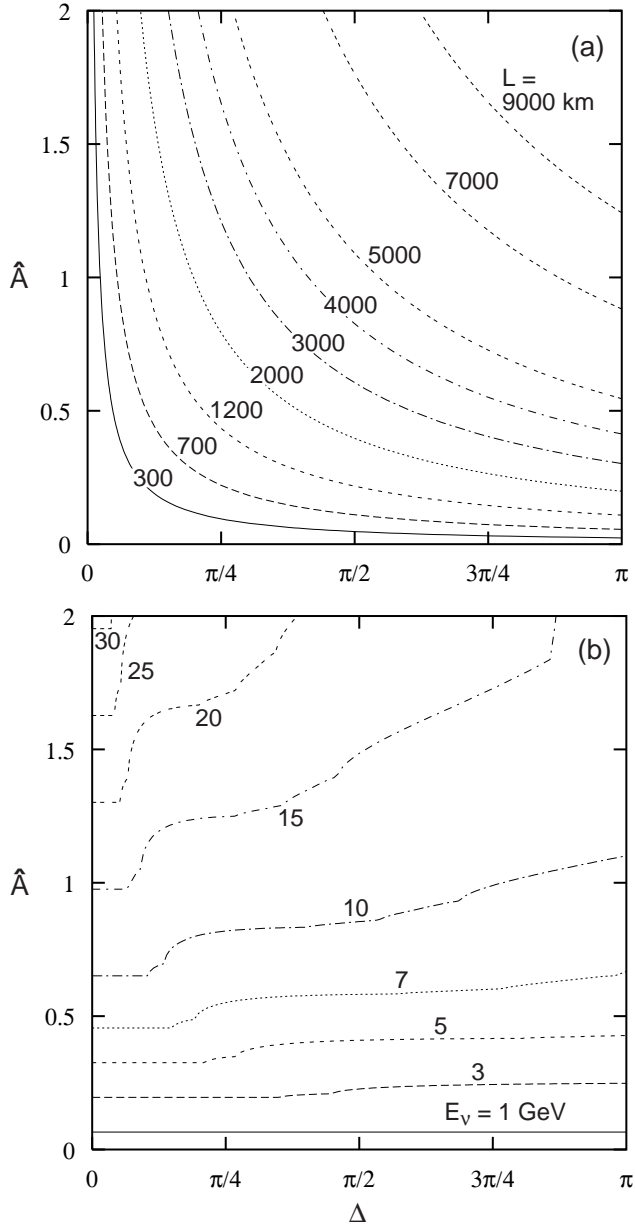,width=3.5in}
\medskip

\caption{Contours of (a) $L$ and (b) $E_\nu$ versus $\hat A$ and
$\Delta$, for $\delta m^2_{31} = 3\times10^{-3}$~eV$^2$.}
\label{fig:LE}
\end{figure}

For reference, the conversion from $\hat A$ and $\Delta$ to $L$ and
$E_\nu$ is shown in Fig.~\ref{fig:LE}. For neutrinos with $\delta
m^2_{31} > 0$ or anti--neutrinos with $\delta m^2_{31} < 0$, $\hat A =
1$ corresponds to an MSW resonance. For neutrinos, it can be shown that
the choice $\hat A = 1/2$ maximizes both the $\sin\delta$ and
$\cos\delta$ terms for a given $\Delta$; for anti-neutrinos the $\hat A$
that maximizes the $\sin\delta$ and $\cos\delta$ terms varies with
$\Delta$.

% 1

We make two observations regarding the approximate probability formulas
above, the consequences of which are discussed below:
\begin{enumerate}

\item[(i)] Both terms that depend on the $CP$ phase $\delta$ vanish when
$g = 0$, i.e., at $\hat A\Delta = n\pi$, where $n$ is an integer. The
$y^2$ term also vanishes in this case, so that only the $x^2$ term
survives.

\item[(ii)] The $\cos\delta$ term vanishes when $\Delta =
(n-{1\over2})\pi$, while the $\sin\delta$ term vanishes when
$\Delta = n \pi$.

\end{enumerate}
The above statements are true for both neutrinos and anti-neutrinos.

The first observation implies that there is no sensitivity to the
$CP$--violating phase $\delta$ when $\langle N_e \rangle L = \int N_e
dL$ is an integer multiple of $\sqrt2 \pi/G_F$, where $\langle N_e
\rangle$ is the average value of $N_e$ for the neutrino path.
Numerically, for $n=1$, this condition is
\begin{equation}
\langle N_e \rangle L \simeq 16275{\rm~km} \,,
\label{eq:Lmat}
\end{equation}
or, for the Earth's density profile,
\begin{equation}
L \simeq 7600{\rm~km} \,.
\label{eq:Lmat2}
\end{equation}
This distance has
a simple physical interpretation: it is the characteristic oscillation
wavelength due to the matter interaction~\cite{matter}. Furthermore, the
condition in Eq.~(\ref{eq:Lmat}) is {\it independent of all oscillation
parameters}. It is also independent of $E_\nu$. It has often been noted
that $CP$ violation is strongly
suppressed in long baseline experiments of order 7300~km (nominally the
distance from Fermilab to Gran Sasso); we see that this is a universal
effect that occurs because $L$ is close to the oscillation length due to
matter.  Furthermore, the term proportional to $y^2$ also vanishes,
which means that there is also no dependence on $\delta m^2_{21}$ or
$\theta_{12}$ at this distance, at least to second order in the small
parameters.  Therefore this distance is especially well--suited for
measuring $\theta_{13}$ without the complications of disentangling it
from $\delta$, $\theta_{12}$, or $\delta m^2_{21}$. For baselines
greater than about 4000~km the constant density approximation loses
accuracy (see results in the Appendix), so that the critical distance in
Eq.~(\ref{eq:Lmat2}) is not exact, but does explain
semi--quantitatively the weakness of $CP$ violating effects near that
distance.

The second observation relates to the relative strength of the
$\sin\delta$ and $\cos\delta$ terms in $P(\nu_\mu \to \nu_e)$.
In short $L$, low $E_\nu$
experiments the matter effects are small and the leading terms of the
oscillation probability are given by the vacuum formulas. Then $L$ and
$E_\nu$ can be chosen such that only the explicitly $CP$--violating
$\sin\delta$ term survives (e.g., when $\Delta = \pi/2$), and $CP$
violation can be measured directly by comparing $P(\nu_\mu \to \nu_e)$
and $P(\bar\nu_\mu \to \bar\nu_e)$ (although even for $L \sim {\rm~few~}
100$~km there are small matter corrections that must be considered).
However, as is evident from Eqs.~(\ref{eq:P}) and (\ref{eq:Pbar}), when
$\theta_{13}$ is small the relative strengths of the
$\sin\delta$ and $\cos\delta$ terms in the presence of large matter
corrections at longer $L$ can be selected by an appropriate choice of
$\Delta$ {\it in exactly the same way} as in the short $L$,
vacuum--like case. That is, the $\delta$ dependence with matter
effects included can be made pure $\sin\delta$ for 
\begin{equation}
L/E_\nu \simeq (2n-1) (410 {\rm~km/GeV~}) \left( {3\times10^{-3}{\rm~eV}^2
\over |\delta m^2_{31}|} \right) \,,
\label{eq:sind}
\end{equation}
where $n$ is an integer~\footnote{The misconception that the
$\cos\delta$ term dominates at large $L$ and $E_\nu$ comes from
extending the large $E_\nu$ approximation beyond its range of validity, as
discussed in Ref.~\cite{freund}.}. The only caveat is that matter
corrections are much larger for longer $L$ and the accuracy of the
determination of $\delta$ may be more subject to knowledge of the
electron density.

%In alternative analytic expressions of the probabilites for oscillations in  
%matter, at arbitrary $\Delta$ the large $E_\nu$ limit gives a
%$\delta$ dependence that comes primarily from the
%$\cos\delta$ term, which is $CP$ conserving. Therefore we conclude that
%at larger $L$ and $E_\nu$
%the $\cos\delta$ term is not necessarily dominant. 

\subsection{Orbits in probability space}
\label{sec:orbits}

We assume that $\sin^2 2\theta_{23}$ and $|\delta m^2_{31}|$ are
well--determined (perhaps at the few percent level or better) by a
$\nu_\mu$ survival or $\nu_\mu \to \nu_\tau$ measurement~\cite{Barger:2001yx}, 
and that
$\theta_{12}$ and $\delta m^2_{21}$ are also well--determined (KamLAND
should be able to measure the parameters of the solar LMA solution to
the few percent level~\cite{kamland}). Then the remaining parameters
to be determined are $\delta$, $\theta_{13}$, and the sign of $\delta
m^2_{31}$ (the sign of $\delta m^2_{21}$ is positive for
LMA).

The usual proposal for testing $CP$ violation in the neutrino sector is
to measure both $\nu_\mu \to \nu_e$ and $\bar\nu_\mu \to \bar\nu_e$
probabilities. As $\delta$ varies for given $\theta_{13}$ and
sgn($\delta m^2_{31}$), an elliptical orbit will be traced in $P$--$\bar
P$ space~\cite{jhfsk,minakata}. The shape of the ellipse is determined
by the relative phases of the terms involving $\delta$. We identify
three possible cases:

\begin{enumerate}

\item[(i)] $\Delta \ne n \pi/2$. In this case, both the $\sin\delta$ and
$\cos\delta$ terms are nonzero and the orbit for fixed $\theta_{13}$ is
an ellipse. Each value of $\delta$ gives a distinct point in $P$--$\bar
P$ space for a given $\theta_{13}$. For $\Delta = (n -
{1\over2}){1\over2}\pi$ the ellipse has the maximum
``fatness''~\cite{minakata}, i.e., it is as close as possible to a
circle given the values of $f$ and $\bar f$.

\item[(ii)] $\Delta = (n -{1\over2}) \pi$, where $n$ is an integer.  In
this case the $\cos\delta$ term vanishes and the orbit ellipse collapses to
a line. If $f \simeq \bar f$ (such as at short $L$ where matter effects
are small), $CP$ violation is measured directly by comparing the $\nu$
and $\bar\nu$ event rates (after correcting for the differences in the
cross sections and initial flux normalization).

\item[(iii)] $\Delta = n \pi$. In this case the $\sin\delta$ term
vanishes, the ellipse collapses to a line, and $CP$ violation
is measured indirectly by parametrically determining the value of
$\delta$ and not by the measurement of a $CP$--violating quantity.

\end{enumerate}

\noindent There will be two ellipses for each $\theta_{13}$, one for each sign
of $\delta m^2_{31}$; they both fall into the same class, i.e., if
the ellipse for $\delta m^2_{31} > 0$ is Case~(ii), the ellipse for
$\delta m^2_{31} < 0$ will also be Case~(ii).

\subsection{$CP$ degeneracy: $(\delta,\theta_{13})$ ambiguity}
\label{sec:delta}

\begin{figure}
\centering\leavevmode
\psfig{file=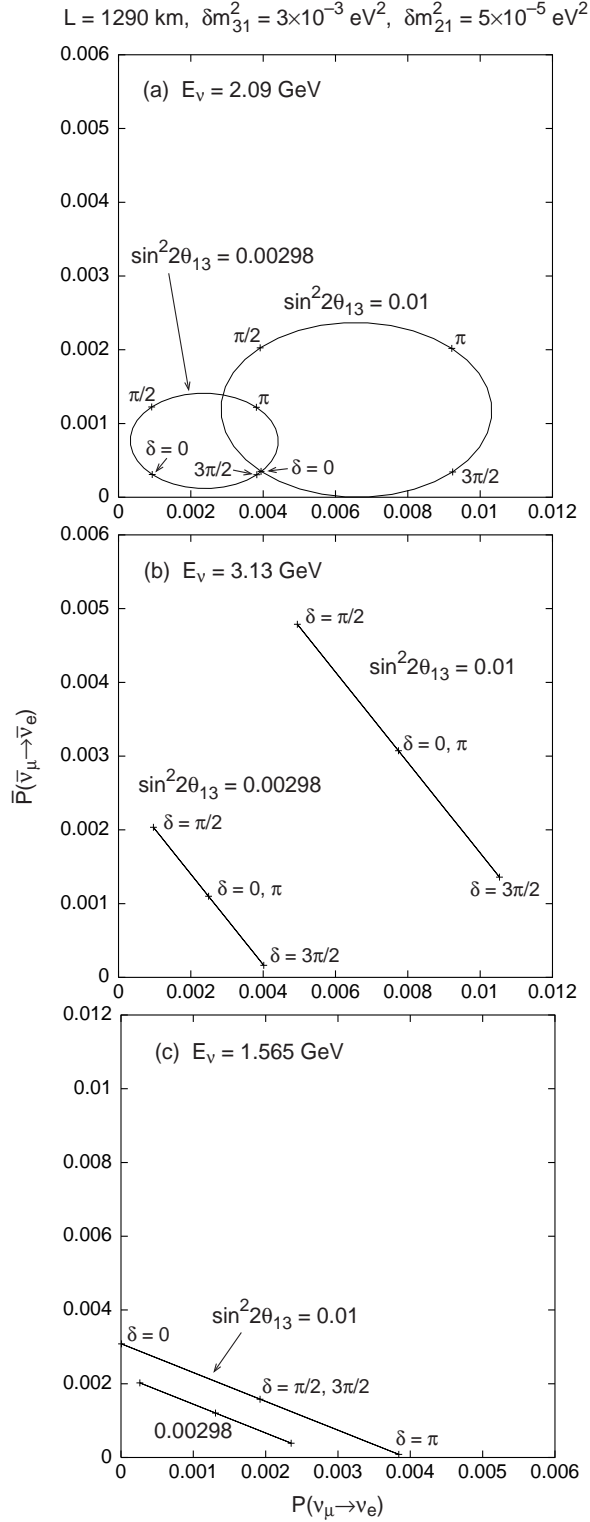,height=8in}
\medskip

\caption{Orbit ellipses showing $(\delta,\theta_{13})$ ambiguity for $L
= 1290$~km with
(a) $E_\nu = 2.09$~GeV ($\Delta = 3\pi/4$),
(b) $E_\nu = 3.13$~GeV ($\Delta = \pi/2$), and
(c) $E_\nu = 1.565$~GeV ($\Delta = \pi$), for $\sin^22\theta_{13} =
0.01$ and $0.00298$. The other parameters are
$\delta m^2_{31} = 3\times10^{-3}$~eV$^2$,
$\delta m^2_{21} = 5\times10^{-5}$~eV$^2$,
$\sin^22\theta_{23} = 1$, and $\sin^22\theta_{12} = 0.8$. The value of
$\delta$ varies around the ellipse.
In (b) and (c) the ellipse collapses to a line and the ambiguity
reduces to a $(\delta, \pi - \delta)$ or $(\delta,2\pi -\delta)$ ambiguity,
respectively, and different values of $\theta_{13}$ do not overlap (for
the same sgn($\delta m^2_{13}$)).}
\label{fig:delta}
\end{figure}

In many cases the parameters $(\delta,\theta_{13})$ can give the same
probabilities as another pair of parameters
$(\delta^\prime,\theta_{13}^\prime)$, for fixed values of the other
oscillation parameters; this is known as the ``$(\delta,\theta_{13})$
ambiguity''~\cite{ambiguity}. Using Eqs.~(\ref{eq:P}) and (\ref{eq:Pbar}),
the general formulas for the parameters $(x^\prime,\delta^\prime)$ that
give the same $P$ and $\bar P$ as $(x,\delta)$ for $\Delta \ne n\pi/2$
(Case~(ii)) are
\begin{eqnarray}
x^\prime \cos\delta^\prime &=&
x\cos\delta + {(f+\bar f)(x^2-x^{\prime 2})\over4yg\cos\Delta} \,,
\label{eq:xcosd}\\
x^\prime \sin\delta^\prime &=&
x\sin\delta - {(f-\bar f)(x^2-x^{\prime 2})\over4yg\sin\Delta} \,.
\label{eq:xsind}
\end{eqnarray}
Equations~(\ref{eq:xcosd}) and
(\ref{eq:xsind}) can be used to derive
\begin{equation}
x^{\prime 2} - x^2 = {4yg\sin2\Delta
\left[yg\sin2\Delta + xf\sin(\Delta-\delta)
+ x\bar f\sin(\Delta+\delta) \right] \over f^2 + \bar f^2
- 2f\bar f\cos2\Delta} \,,
\label{eq:x2}
\end{equation}
from which $\delta^\prime$ can then be determined from
Eq.~(\ref{eq:xcosd}) or (\ref{eq:xsind}). In particular, a
set of parameters which violates $CP$ ($\sin\delta^\prime \ne 0$) can
be degenerate with another set of parameters, with a different
$\theta_{13}$, that conserves $CP$ ($\sin\delta = 0$). It can be shown
that in all cases real solutions exist for $x^\prime$, so there will be
an ambiguity between two sets of oscillation parameters if
$|\sin\delta^\prime| \le 1$. Therefore we conclude that the use  of a
monoenergetic beam at a fixed $L$ will necessarily entail parameter
ambiguities if only the channels $\nu_\mu \to \nu_e$ and $\bar\nu_\mu \to
\bar\nu_e$ are measured. An example is shown in Fig.~\ref{fig:delta}a
for $\Delta =3\pi/4$.

When $\Delta \ne n\pi/2$, the $(\delta,\theta_{13})$ ambiguity can give
a degeneracy between $CP$ violating ($CPV$) and $CP$ conserving ($CPC$)
solutions. If $\sin\delta = 0$ in Eq.~(\ref{eq:xsind}), then
$\sin\delta^\prime$ is not zero if $f \ne \bar f$; the difference can be
large if $f$ and $\bar f$ differ substantially due to large matter
effects. For example, in Fig.~\ref{fig:delta}a the prediction for
$(P,\bar P)$ for $(\sin^22\theta_{13},\delta) = (0.01, 0)$ is identical
to that for $(0.00298,1.48\pi)$.

For the Cases~(ii) and (iii) above, the ellipse collapses to a line. The
ambiguities then reduce to $x^\prime = x$ (see Eq.~(\ref{eq:x2})) and
$\sin\delta^\prime = \sin\delta$ in Case~(ii) and $\cos\delta^\prime =
\cos\delta$ in Case~(iii). Thus the ambiguity no longer involves
$\theta_{13}$ (and hence in principle $\theta_{13}$ is determined, at
least as far as the $(\delta,\theta_{13})$ ambiguity is concerned), but
instead is a $(\delta,\pi-\delta)$ ambiguity (which does not mix $CPC$
and $CPV$ solutions) in Case~(ii) and a $(\delta,2\pi-\delta)$ ambiguity in
Case~(iii). Examples for Cases~(ii) and (iii) are shown in
Figs.~\ref{fig:delta}b and \ref{fig:delta}c, respectively. For $\Delta
\simeq n\pi/2$ the orbit ellipse is very skinny and the ambiguous
$\theta_{13}$ values are close to each other, which qualitatively is
similar to either Case~(ii) or Case~(iii).

Note that in both Figs.~\ref{fig:delta}b and \ref{fig:delta}c the orbit
line has a negative slope. It can be shown that for $\hat A < 1$
(i.e., density less than the critical density for resonance) the
orbit lines in ($P,\bar P$) space have negative slope for
$\Delta = n\pi/2$. For $\hat A > 1$, the orbit lines for $\Delta = n\pi/2$
have positive slope.

% 2

\subsection{Mass hierarchy degeneracy: sgn($\delta m^2_{31}$) ambiguity}
\label{sec:sign}

\begin{figure}
\centering\leavevmode
\psfig{file=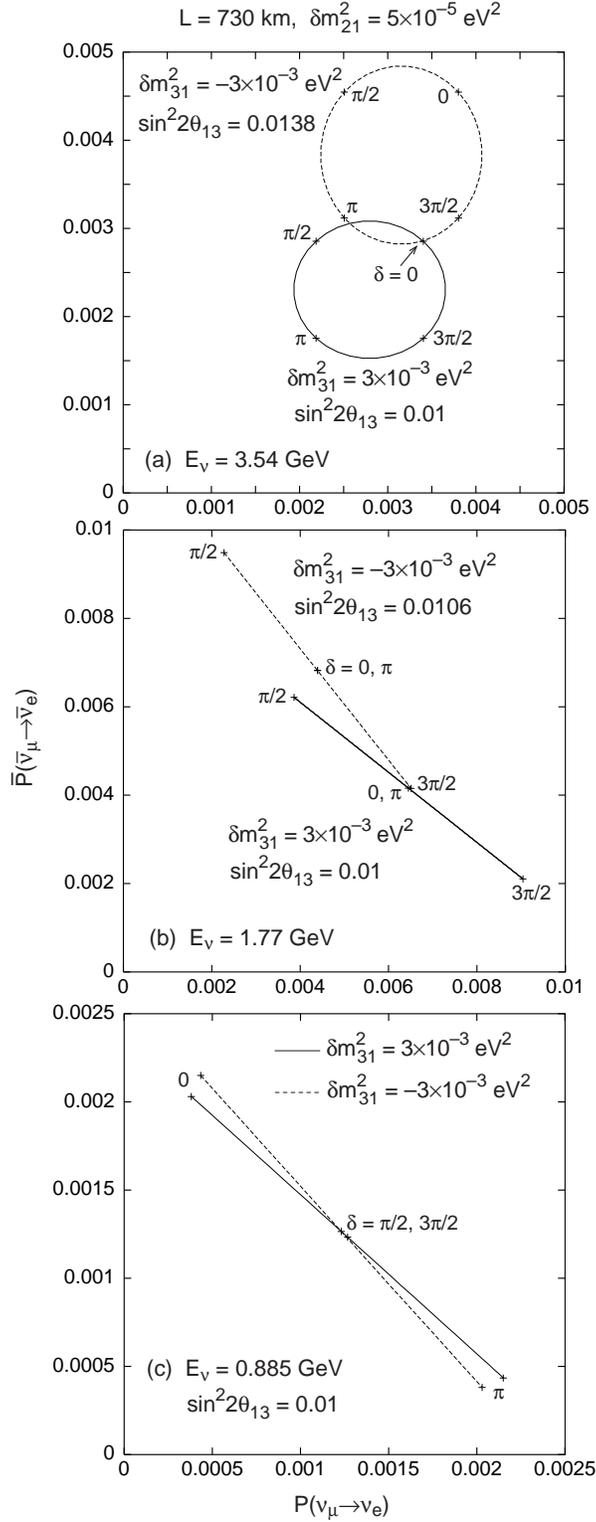,height=8in}
\medskip

\caption{Sgn($\delta m^2_{31}$) ambiguity for $L = 730$~km with
(a) $E_\nu = 3.54$~GeV ($\Delta = \pi/4$),
(b) $E_\nu = 1.77$~GeV ($\Delta = \pi/2$), and
(c) $E_\nu = 0.885$~GeV ($\Delta = \pi$).
The other parameters are $\delta m^2_{21} = 5\times10^{-5}$~eV$^2$,
$\sin^22\theta_{23} = 1$, and $\sin^22\theta_{12} = 0.8$.}
\label{fig:sign}
\end{figure}

In addition to the ($\delta, \theta_{13}$) ambiguity discussed above 
for a given sgn($\delta m^2_{31}$), in some cases there are also
parameters ($\delta^\prime$, $\theta_{13}^\prime$) with $\delta m^2_{31}
< 0$ that give the same $P$ and $\bar P$ as ($\delta$, $\theta_{13}$)
with $\delta m^2_{31}>0$. Three examples of the sgn($\delta m^2_{31}$)
ambiguity are shown in Fig~\ref{fig:sign}. Furthermore, there is also a
($\delta^\prime,\theta_{13}^\prime$) ambiguity for $\delta m^2_{31} <
0$, so in principle there can be a four--fold ambiguity, i.e., four sets
of $\delta$ and $\theta_{13}$ (two for $\delta m^2_{31} > 0$ and two for
$\delta m^2_{31} < 0$) that give the same $P$ and $\bar P$.

As with the $(\delta,\theta_{13})$ ambiguity, the sgn($\delta m^2_{31}$)
ambiguity can mix $CP$ conserving and $CP$ violating solutions. For
example, in Fig.~\ref{fig:sign}a the prediction for $(P,\bar P)$ for
$(\sin^22\theta_{13},\delta) = (0.01, 0)$ with $\delta m^2_{31} =
3\times10^{-3}$~eV$^2$ is identical to that for $(0.0138,4\pi/3)$
with $\delta m^2_{31} = - 3\times10^{-3}$~eV$^2$.

Although the general equations for the sgn($\delta m^2_{31}$) ambiguity
are somewhat messy, for the case $\Delta = (n-{1\over2})\pi$ the values
of $(x^\prime,\delta^\prime)$ for $\delta m^2_{31} < 0$ that give the
same $P$ and $\bar P$ as $(x,\delta)$ for $\delta m^2_{31} > 0$ are
determined by
\begin{eqnarray}
x^{\prime2} &=&
{x^2(f^2+\bar f^2-f\bar f)-2yg(f-\bar f)x\sin\delta\sin\Delta
\over f\bar f} \,,
\label{eq:x22}\\
x^\prime\sin\delta^\prime &=&
x\sin\delta {f^2+\bar f^2-f\bar f\over f\bar f}
- {x^2\over\sin\Delta}{f^2+\bar f^2\over f\bar f}{f-\bar f\over 2yg} \,.
\label{eq:xsind2}
\end{eqnarray}
If $\sin\delta = 0$ then Eq.~(\ref{eq:xsind2}) reduces to
\begin{equation}
\sin\delta^\prime =
- x {f^2+\bar f^2\over f\bar f} {f - \bar f\over 2yg\sin\Delta}
\sqrt{f\bar f\over f^2+\bar f^2 - f\bar f} \,,
\label{eq:sindprime}
\end{equation}
which is not zero if $f\ne\bar f$, i.e., whenever there are matter
effects, so there is a potential $CPC/CPV$ confusion as long as
the right--hand side of Eq.~(\ref{eq:sindprime}) has magnitude less than
unity. It is possible to have $\delta^\prime = \pi/2$ when $\delta = 0$,
i.e., $CPC$ can be confused with maximal $CPV$ (see
Fig.~\ref{fig:sign}b).

The ambiguity between parameters with $\delta m^2_{31} > 0$ and $\delta
m^2_{31} < 0$ occurs only for some values of $\delta$, and does not
occur at all if matter effects are large enough (i.e., $L$ and
$\theta_{13}$ are large enough)~\cite{Barger:2000cp,Lipari:1999wy}.  The
conditions for the existence of this ambiguity will be discussed further
in Sec.~\ref{sec:resolving2}.  Note, however, that the sgn($\delta
m^2_{31}$) ambiguity can still confuse different values of $\delta$ {\it
and} $\theta_{13}$ even for $\Delta = n\pi/2$ (see, e.g.,
Figs.~\ref{fig:sign}b and \ref{fig:sign}c), unlike the ($\delta,
\theta_{13}$) ambiguity where $\theta_{13}$ is removed from the
ambiguity for $\Delta = n\pi/2$.

%so that eliminating the
%$(\delta,\theta_{13})$ ambiguity does not necessarily eliminate the
%sgn($\delta m^2_{13}$) ambiguity.

% 3

\subsection{Atmospheric angle degeneracy:
($\theta_{23}\,,\pi/2-\theta_{23}$) ambiguity}
\label{sec:theta23}

\begin{figure}
\centering\leavevmode
\psfig{file=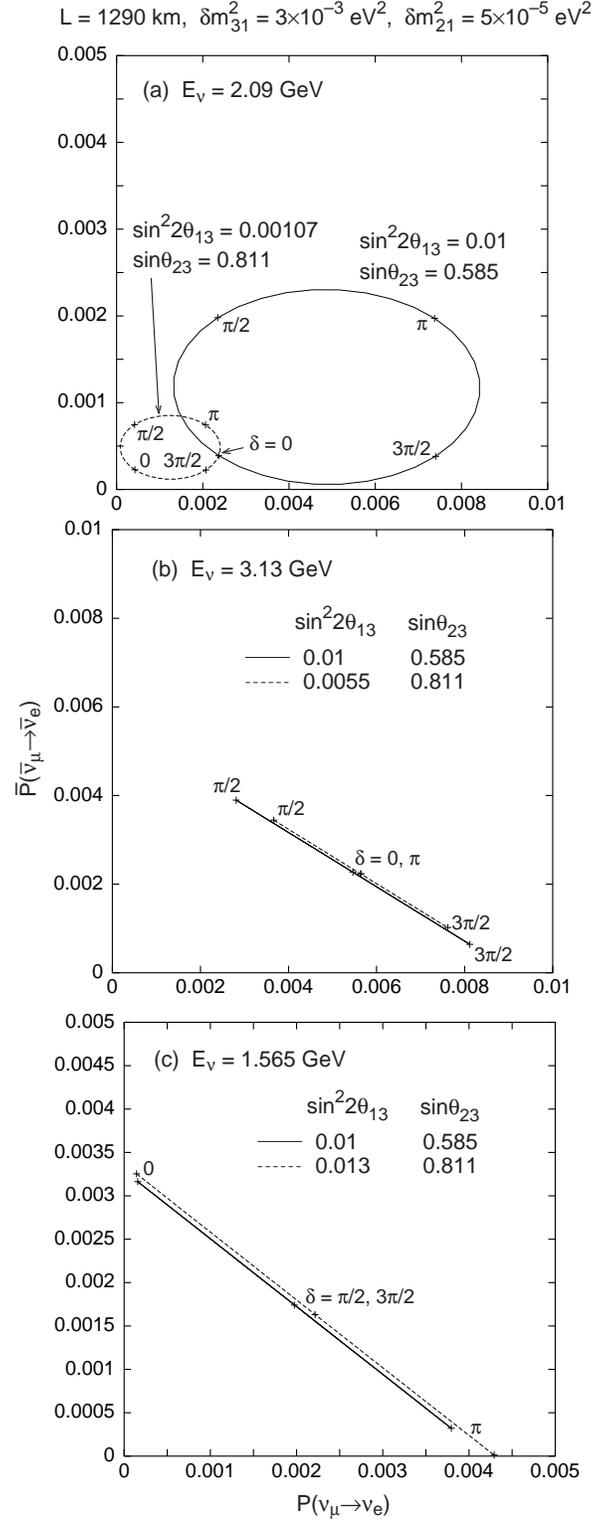,height=8in}
\medskip

\caption{($\theta_{23}\,,\pi/2-\theta_{23}$) ambiguity for $L = 1290$~km with
(a) $E_\nu = 2.09$~GeV ($\Delta = 3\pi/4$),
(b) $E_\nu = 3.13$~GeV ($\Delta = \pi/2$), and
(c) $E_\nu = 1.565$~GeV ($\Delta = \pi$).
The other parameters are $\delta m^2_{21} = 5\times10^{-5}$~eV$^2$,
and $\sin^22\theta_{12} = 0.8$.}
\label{fig:theta23}
\end{figure}

There is yet another ambiguity in the determination of $\delta$ and
$\theta_{13}$, which involves the value of $\theta_{23}$. In practice it
is only $\sin^22\theta_{23}$ that is determined by a $\nu_\mu$ survival
measurement (for now we ignore matter corrections to $\nu_\mu \to
\nu_\mu$, which are relatively small for oscillations involving active
flavors), so $\theta_{23}$ cannot be distinguished from $\pi/2 -
\theta_{23}$. The effect of this degeneracy can be seen by
interchanging $\sin\theta_{23}$ and $\cos\theta_{23}$ in
Eqs.~(\ref{eq:x}) and (\ref{eq:y}). For $\theta_{23} \simeq \pi/4$ (the
favored solution from atmospheric data) the ambiguity vanishes, but for
$\sin^22\theta_{23} \simeq 0.9$ it can have a sizable effect, since in
this case $\sin^2\theta_{23}= 0.35$ and $\cos^2\theta_{23}=0.65$. Three
examples of the $\theta_{23}$ ambiguity are shown in
Fig.~\ref{fig:theta23}. The $\theta_{23}$ ambiguity can also mix $CPC$
and $CPV$ solutions; for example, in Fig.~\ref{fig:theta23}a, the
prediction for $(P,\bar P)$ for $(\sin^22\theta_{13},\sin\theta_{23},
\delta) = (0.01, 0.585, 0)$ is identical to that for $(0.00107, 0.811,
4\pi/3)$.

% 4

As with the sgn($\delta m^2_{31}$) ambiguity, the equations for the
$\theta_{23}$ ambiguity are rather messy in the general case. For the
special case $\Delta = (n-{1\over2})\pi$, we have
\begin{eqnarray}
\sin^22\theta_{13}^{\prime} &=&
\sin^22\theta_{13} \tan^2\theta_{23}
+ {\alpha^2 g^2 \sin^22\theta_{12} \over f\bar f}(1-\tan^2\theta_{23}) \,,
\label{eq:x23}\\
\sin2\theta_{13}^\prime \sin\delta^\prime &=&
\sin2\theta_{13} \sin\delta
+ {\alpha g(f-\bar f)\sin2\theta_{12} \over f\bar f}
{\cot2\theta_{23} \over\sin\Delta} \,,
\label{eq:xsind3}
\end{eqnarray}
where ($\delta, \theta_{13}$) are the parameters that give a certain
$(P,\bar P)$ for $0 < \theta_{23} < \pi/4$ and ($\delta^\prime,
\theta_{13}^\prime$) are the parameters that give the same $(P,\bar P)$
for $\pi/2 - \theta_{23}$.  We see that even for $\Delta =
n\pi/2$ the $\theta_{23}$ ambiguity can mix $CPC$ and $CPV$
solutions, since $\sin\delta =0$ does not necessarily imply
$\sin\delta^\prime = 0$. Furthermore, even for $\Delta =
n\pi/2$ the $\theta_{23}$ ambiguity mixes solutions with
different $\theta_{13}$ (see Eqs.~(\ref{eq:x23}) and (\ref{eq:xsind3}),
and Figs.~\ref{fig:theta23}b and \ref{fig:theta23}c),
unlike the $(\delta, \theta_{13})$ ambiguity where $\theta_{13}$ is
removed from the ambiguity for $\Delta = n\pi/2$.

Since $\alpha$ is a small parameter (and possibly even small compared to
$\sin2\theta_{13}$), the numerical uncertainty in $\delta$ due to the
$\theta_{23}$ ambiguity is generally small, of order $0.07\pi$ or less,
when $\Delta = (n - {1\over2})\pi$, $L = 2900$~km, $\delta m^2_{21} =
10^{-4}$~eV$^2$, and $\sin^22\theta_{13} = 0.01$. This effect is of
order the expected experimental uncertainty in
$\delta$~\cite{jhfsk,peak}. The size of the $\delta$ ambiguity decreases
with decreasing matter effect (smaller $L$) and with decreasing $\delta
m^2_{21}$, so for a wide range of parameters the $CPC/CPV$ confusion
from the $\theta_{23}$ ambiguity is not too severe. On the other hand,
the $\sin^22\theta_{13}$ confusion is approximately a factor
$\tan^2\theta_{23}$ (see Eq.~(\ref{eq:x23})), which lies roughly in the
range ${1\over2}$ to 2 for $\sin^22\theta_{23} \ge 0.9$.

\subsection{Comments on parameter degeneracies}
\label{sec:comments}

In the preceding three sections we have shown that in principle there
can be as much as an eight--fold ambiguity in determining $\delta$ and
$\theta_{13}$ from $P(\nu_\mu \to \nu_e)$ and ${\bar P}(\bar\nu_\mu \to
\bar\nu_e)$ at a single $L$ and $E_\nu$, which comes from the presence
of three independent two--fold ambiguities: $(\delta,\theta_{13})$,
sgn($\delta m^2_{31}$), and ($\theta_{23}\,,\pi/2-\theta_{23}$).  For
each type of ambiguity there is the possibility of being unable to
distinguish between $CP$ violating and $CP$ conserving
parameters. Measurements at multiple $L$ and $E_\nu$ can be used to help
discriminate the different degenerate solutions, but that would involve
extra detectors or a much longer total running time, and probably
reduced statistics for each $(L,E_\nu)$ combination. In the next section
we will explore what $L$ and $E_\nu$ values do best at resolving these
potential degeneracies without resorting to measurements at different
$L$ and/or $E_\nu$. We then will examine what $L$ and $E_\nu$ for a
second measurement can remove the remaining degeneracies.

In the Appendix we demonstrate that the analytic expressions are accurate
for $E_\nu > 0.5$ GeV for baselines up to 4000-5000 km. For much lower
$E_\nu$ (as low as 0.05~GeV) they are still accurate at shorter
distances ($L \lsim 350$~km) if $\alpha$ and $\theta_{13}$ are not too
large (see the discussion in the Appendix). Therefore we expect that the
qualitative aspects of the three ambiguities are unchanged for
short $L$, low $E_\nu$ experiments such as CERN--Frejus.

%In principle there is also yet another ambiguity in the sign of $\delta
%m^2_{21}$, which will affect the probabilities ($y \to -y$ in the
%approximate probability formulas) and lead to another two--fold
%ambiguity in the determination of $\delta$ and $\theta_{13}$.  KamLAND
%will only measure $\sin^22\theta_{12}$ and hence cannot resolve this
%ambiguity [right?]. On the other hand, if we believe that the solar
%solution is LMA, then we must have $\delta m^2_{21} > 0$. The point here
%is that measurements of $P(\nu_\mu \to \nu_e)$ and $P(\bar\nu_\mu \to
%\bar\nu_e)$ at a single $L$ and $E_\nu$ will not provide an independent
%confirmation of sgn($\delta m^2_{21}$) [I think this is true].

\section{Resolving parameter degeneracies}
\label{sec:resolving}

\subsection{Resolving the ($\delta,\theta_{13}$) ambiguity}
\label{sec:resolving1}

As discussed in Secs.~\ref{sec:orbits} and \ref{sec:delta}, the choice
$\Delta = n \pi/2$ causes the orbit ellipse in $(P,\bar P)$ space to
collapse to a line and the $(\delta,\theta_{13})$ ambiguity reduces to
one involving only $\delta$, i.e., the combination of $P$ and $\bar P$
gives a unique value of $\theta_{13}$ (at least for one sign of $\delta
m^2_{31}$). Furthermore, since $\delta$ only becomes confused with
$\pi-\delta$ (in Case~(ii) of Sec.~\ref{sec:delta}) or $2\pi-\delta$ (in
Case~(iii)), $CP$ conserving solutions never become mixed with $CP$
violating ones. Case~(ii) (with $\Delta = (n-{1\over2})\pi$) has another
advantage in that the $\nu_\mu \to \nu_\tau$ oscillation is
approximately maximal (see Eq.~(\ref{eq:Pmutau})), which would
facilitate a better measurement of $\sin^2 2\theta_{23}$ and $\delta
m^2_{31}$. Therefore the choice $\Delta = (n-{1\over2})\pi$ is the best
for resolving the $(\delta,\theta_{13})$ ambiguity. Some representative
beam energies for particular baselines are given in
Table~\ref{tab:Delta}.

\begin{table}[t]
%\squeezetable
\caption[]{Possible neutrino beam energies $E_\nu$ (in GeV) versus
baseline (in km) and $\Delta$ that will convert the
$(\delta,\theta_{13})$ ambiguity to a simple $(\delta, \pi-\delta)$
ambiguity, for $\delta m^2_{31} = 3\times10^{-3}$~eV$^2$. For other
values of $\delta m^2_{31}$, $E_\nu$ scales proportionately with
$\delta m^2_{31}$. Only values of $E_\nu > 0.5$~GeV are considered.}
\begin{tabular}{c|ccccccc}
$\Delta$ & 300~km & 730~km & 1290~km & 1770~km & 2100~km & 2600~km & 2900~km \\
\\
${\pi\over2}$   & 0.73 & 1.77 & 3.13 & 4.29 & 5.12 & 6.34 & 7.03\\
${3\pi\over2}$  &      & 0.59 & 1.04 & 1.43 & 1.71 & 2.11 & 2.34\\
${5\pi\over2}$  &      &      & 0.63 & 0.86 & 1.02 & 1.27 & 1.41\\
${7\pi\over2}$  &      &      &      & 0.61 & 0.73 & 0.91 & 1.00\\
${9\pi\over2}$  &      &      &      &      & 0.57 & 0.70 & 0.78\\
\end{tabular}
\label{tab:Delta}
\end{table}

\subsection{Resolving the sgn($\delta m^2_{31}$) ambiguity}
\label{sec:resolving2}

The parameter degeneracy associated with the sign of $\delta m^2_{31}$
can be overcome if there is a large matter effect that splits $P$ and
$\bar P$, e.g., if $L$ is sufficiently long and $\theta_{13}$ is not too
small~\cite{Barger:2000cp,Lipari:1999wy}. 
To determine the minimum value of $\theta_{13}$ that avoids the
sgn($\delta m^2_{31}$) ambiguity, we must first find the region in
$(P,\bar P)$ space covered by each sgn($\delta m^2_{31}$), and then
determine the condition on $\theta_{13}$ that ensures the regions for
different sgn($\delta m^2_{31}$) do not overlap.

\begin{figure}[h!]
\centering\leavevmode
\psfig{file=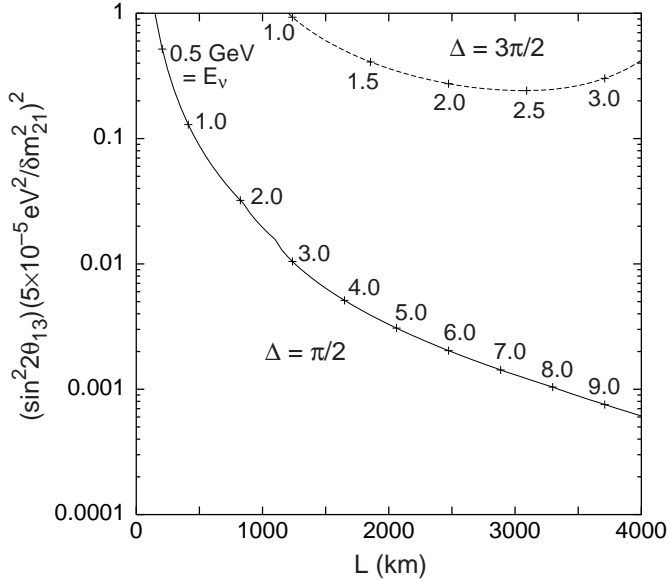,width=3.5in}
\medskip

\caption{Minimum value of $\sin^22\theta_{13}/(\delta m^2_{21})^2$
that avoids the sgn($\delta m^2_{31}$) ambiguity, plotted versus
$L$ for $\Delta = \pi/2$~(solid curve) and $3\pi/2$ (dashed), with
$\delta m^2_{31} = 3\times10^{-3}$~eV$^2$. The corresponding values of
$E_\nu$ are marked on the curves.}
\label{fig:minx}
\end{figure}

The orbit ellipse for a given sgn($\delta m^2_{31}$) moves as
$\theta_{13}$ changes, sweeping out a region in $(P,\bar P)$ space. All
points on each orbit ellipse (for a given $\theta_{13}$) that lie inside
the region will overlap an orbit ellipse for a different $\theta_{13}$
(this is what leads to the ($\delta, \theta_{13}$) ambiguity).
However, the points on the orbit ellipse that lie on the boundaries of
the region do not have a ($\delta,\theta_{13}$) ambiguity, i.e.,
there are unique values of $\theta_{13}$ and $\delta$ for that point.
This implies that for points on the boundary of the region, $x =
x^\prime$ and $\delta = \delta^\prime$ in
Eqs.~(\ref{eq:xcosd})--(\ref{eq:x2}). For $\Delta \ne n\pi/2$, the
condition becomes $xf\sin(\Delta - \delta) + x\bar f\sin(\Delta +
\delta) + yg\sin2\Delta = 0$. Solving for $\delta$ and substituting into
Eqs.~(\ref{eq:P}) and (\ref{eq:Pbar}) we find the coordinates of the
$\delta m^2_{31} > 0$ envelope in $(P,\bar P)$ space are given by
\begin{equation}
P = x^2 f^2 + y^2 g^2 - {2 y^2 g^2 f^2 \sin^22\Delta \pm 2ygf\sqrt{z}
(f \cos2\Delta - \bar f)\over f^2 + \bar f^2 - 2 f\bar f\cos2\Delta} \,,
\label{eq:Penv}
\end{equation}
where
\begin{equation}
z = x^2 (f^2 + \bar f^2 - 2 f\bar f\cos2\Delta) - y^2 g^2 \sin^22\Delta
\,,
\label{eq:z}
\end{equation}
and $\bar P$ is found by interchanging $f \leftrightarrow \bar f$ and
letting $g \to -g$. For $\delta m^2_{31} < 0$, the values of $P$ and
$\bar P$ on the envelope can be found by interchanging $P$ and $\bar P$.
Although the general solution is complicated, for the special case
$\Delta = (n-{1\over2}) \pi$ it is not hard to show that the two
sgn($\delta m^2_{31}$) regions do not overlap if
\begin{equation}
x > {2yg\over f - \bar f} \,.
\label{eq:xlim2}
\end{equation}
Note that matter effects split $f$ and $\bar f$, which decreases the
minimum value of value of $x$ (and hence of $\sin^22\theta_{13}$) needed
to avoid any sgn($\delta m^2_{31}$) ambiguity. Because $x \propto
\sin2\theta_{13}$ and $y \propto \delta m^2_{21}$, the corresponding
minimum value of $\sin^22\theta_{13}$ increases as the square of $\delta
m^2_{21}$. The minimum values of
$\sin^22\theta_{13}/(\delta m^2_{21})^2$ are plotted versus $E_\nu$ for
various values of $L$ in Fig.~\ref{fig:minx}. For $\Delta = \pi/2$, the
sgn($\delta m^2_{31}$) ambiguity would be resolved for
$\sin^22\theta_{13} > 0.01$ (0.04) at $L=1290$~km, the distance from
Fermilab to Homestake, if $\delta m^2_{21} =
5\times10^{-5}~(10^{-4})$~eV$^2$. For $L \simeq 2600$~km
(Brookhaven--Homestake or Fermilab--San Jacinto), sgn($\delta m^2_{31}$)
can be determined for values of $\sin^22\theta_{13}$ as low as 0.002
(0.008) for $\delta m^2_{21} = 5\times10^{-5}~(10^{-4})$~eV$^2$.

% 5

Figure~\ref{fig:minx} shows that $\Delta = 3\pi/2$ would be
unsatisfactory in distinguishing sgn($\delta m^2_{31}$); 
in fact, measurements at 
$\Delta=(n-{1\over2})\pi$ do increasingly worse as $n$ increases, 
as can be shown using Eq.~(\ref{eq:f}). For $\Delta =
(n-{1\over2})\pi$, we have $|f/\bar f| = (1+\hat A)/(1-\hat A)$; since
$\hat A$ is proportional to $E_\nu$, and $E_\nu$ decreases with $n$ for
fixed $L$, larger values of $n$ will have smaller $\hat A$. Thus, the
values of $f$ and $\bar f$ will be closer for larger $n$, reducing the
size of the matter effect (at least as far as splitting $P$ and $\bar P$
is concerned). For $\Delta = 3\pi/2$ and  $\delta m^2_{32} =
5\times10^{-5}$~eV$^2$, the value of $\sin^22\theta_{13}$ must be
greater than about 0.25 which is excluded by CHOOZ~\cite{CHOOZ}. Even
the most optimistic case for $\Delta = 3\pi/2$ (which occurs for the
highest value of $\delta m^2_{21} (\simeq 10^{-4}$~eV$^2)$ allowed in the
LMA region) requires $\sin^22\theta_{13} \gsim 0.06$.
 Thus, the proposal of Ref.~\cite{Marciano:2001tz} 
to perform experiments at higher $n$ suffers from an inability to
determine sgn($\delta m^2_{31}$).  
Practically speaking, only $n=1$ will provide sufficient discrimination
for sgn($\delta m^2_{31}$) if $\Delta$ is restricted to the values
$(n-{1\over2})\pi$.
We henceforth restrict ourselves to this case.

\begin{figure}
\centering\leavevmode
\psfig{file=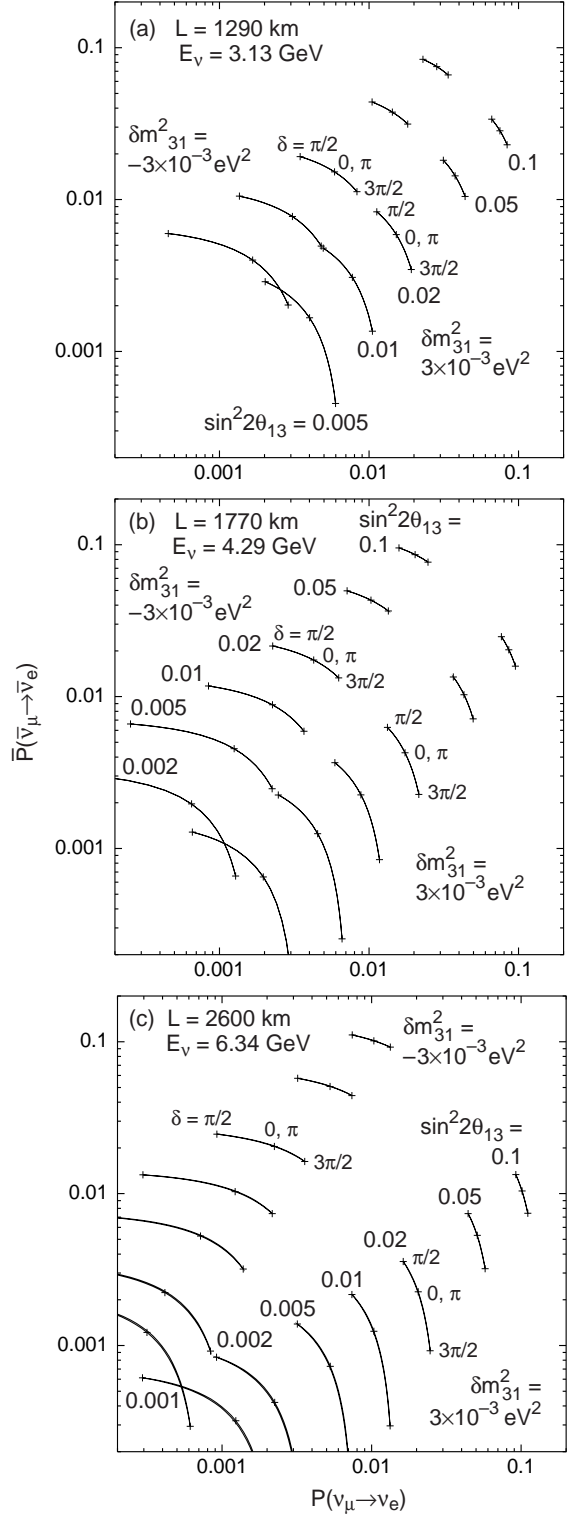,height=8in}
\medskip

\caption{Resolution of combined $(\delta,\theta_{13})$ and sgn($\delta
m^2_{31}$) ambiguities when $\Delta = \pi/2$, for (a) $L = 1290$~km,
(b) $L = 1770$~km, and (c) $L = 2900$~km, with $|\delta m^2_{31}| =
3\times10^{-3}$~eV$^2$, $|\delta m^2_{21}| = 5\times10^{-5}$~eV$^2$, and
$\sin^22\theta_{23} = 1$.}
\label{fig:deltasign1}
\end{figure}

\begin{figure}
\centering\leavevmode
\psfig{file=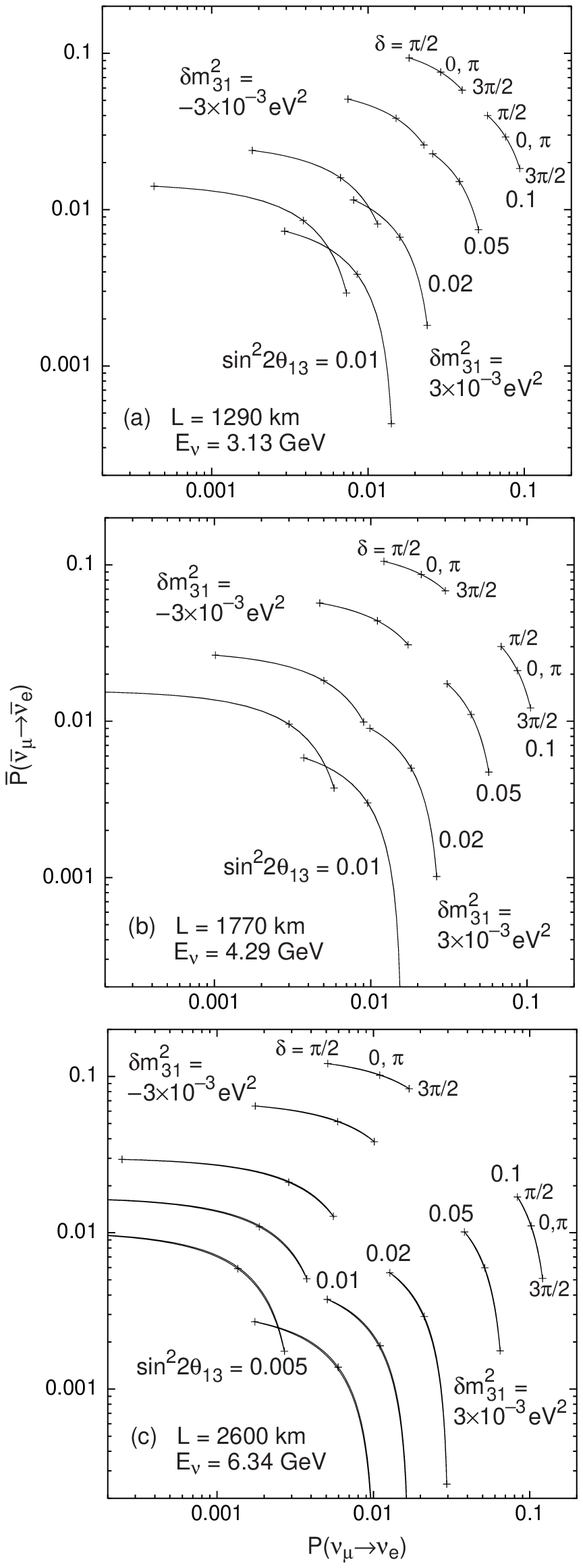,height=8in}
\medskip

\caption{Same as Fig.~\ref{fig:deltasign1}, except for
$|\delta m^2_{21}| = 10^{-4}$~eV$^2$.}
\label{fig:deltasign2}
\end{figure}

By combining $\Delta = \pi/2$ with a sufficiently long $L$, the combined
four-fold ambiguity involving $\delta$, $\theta_{13}$, and sgn($\delta
m^2_{31}$) can be reduced to a simple ($\delta,\pi-\delta$) ambiguity
that in principle determines whether $CP$ is conserved or violated. Some
possibilities are shown in Figs.~\ref{fig:deltasign1} and
\ref{fig:deltasign2}. As Fig.~\ref{fig:deltasign1} shows, for $\delta
m^2_{21} = 5\times10^{-5}$~eV$^2$, $L = 1290$~km is (barely) sufficient
to distinguish sgn($\delta m^2_{31}$) for $\sin^22\theta_{13}$ as low as
0.01. However, for $\delta m^2_{21} = 10^{-4}$~eV$^2$, $L > 2000$~km is
needed. In practice, experimental uncertainties and uncertainties
on the matter distribution~\cite{Geller:2001ix} increase the
likelihood of having a sgn($\delta m^2_{31}$) ambiguity, so that a separation
of the two regions greater than the size of the experimental
uncertainties is required.

% 6

% 7

\subsection{Resolving the ($\theta_{23}\,,\pi/2-\theta_{23}$) ambiguity}
\label{sec:resolving3}

Even if $\Delta$ is chosen to mitigate the effects of the
$(\delta,\theta_{13})$ ambiguity, and $L$ is chosen long enough to
eliminate the sgn($\delta m^2_{31}$) ambiguity, there still remains the
$(\theta_{23}, \pi/2-\theta_{23})$ ambiguity. If $\theta_{23} \simeq
\pi/4$ this ambiguity disappears, and choosing $L$ and $E_\nu$ such that
$\Delta = \pi/2$ leaves a simple $(\delta,\pi-\delta)$ ambiguity.  If
$\theta_{23}$ deviates from $\pi/4$, then there does not appear to be a
judicious choice of a single $L$ and $E_\nu$ that can resolve the
$\theta_{23}$ ambiguity.

The problem in resolving the $\theta_{23}$ ambiguity lies in the fact
that in the leading term in $P(\nu_\mu \to \nu_e)$ and ${\bar
P}(\bar\nu_\mu \to \bar\nu_e)$, $\sin2\theta_{13}$ is always paired with
$\sin\theta_{23}$ (see Eqs.~(\ref{eq:P}) and (\ref{eq:Pbar})), and so if
there are two values of $\theta_{23}$ derived from the measured value of
$\sin^22\theta_{23}$, there will be two corresponding values of
$\sin^22\theta_{13}$. Since $\sin^22\theta_{23}$ can be as low as 0.9,
the two values of $\sin^2\theta_{23}$ can be as far apart as 0.35 and
0.65, and therefore the ambiguity in $\sin^22\theta_{13}$ can be as
large as a factor 1.86 at leading order (Eq.~(\ref{eq:x23}) in the limit
that $\alpha$ is small). The next--to--leading term in the probabilities
in Eqs.~(\ref{eq:P}) and (\ref{eq:Pbar}) is proportional to
$\sin2\theta_{23}$, and therefore cannot resolve the ambiguity. The last
term in $P(\nu_\mu \to \nu_e)$ is proportional to $\cos^2\theta_{23}$,
so that the relative weighting of the last term compared to the leading
term is affected by the value of $\sin\theta_{23}$. However, the last
term is suppressed by $\alpha^2$, and is generally much smaller than the
leading term (at least for $\sin^22\theta_{13} \ge 0.01$, the
approximate region where superbeam experiments will be able to probe).
Hence, even measurements at a second $L$ and $E_\nu$ would likely be
unable to resolve the $\theta_{23}$ ambiguity if it exists (i.e.,
$\theta_{23}$ not close to $\pi/4$).

If one could also measure $P(\nu_e \to \nu_\tau)$ (see Appendix A for
an approximate analytic expression), then a comparison with $P(\nu_e
\to \nu_\mu)$ should determine whether $\theta_{23}$ is above or below
$\pi/4$; the leading term in $P(\nu_e \to \nu_\tau)$ can be obtained from
the leading term in  $P(\nu_\mu \to \nu_e)$ by the replacement of
$\sin \theta_{23}$ by $\cos \theta_{23}$. 
A $\nu_e\to\nu_\tau$ measurement could be done in a neutrino
factory; in fact, a neutrino factory may be the only practical way to
resolve the $\theta_{23}$ ambiguity, if it exists. A neutrino factory
experiment also provides energy spectrum information that could be
helpful in resolving parameter ambiguities~\cite{ambiguity,nufactamb}.

\subsection{Measurements at a second $L$ and/or $E_\nu$}

As we have demonstrated, measurements at a single $L$ and $E_\nu$ cannot
resolve all parameter ambiguities. A second experiment at a different
$L$ and/or $E_\nu$, with a different value of $\Delta$, is required for
this purpose. The best sets of $L$ and $E_\nu$ are those that are
complementary, i.e., the second experiment should provide the clearest
distinction between the parameter ambiguities of the first
experiment. In this section we discuss three possible scenarios, each
with measurements at two $L$ and $E_\nu$ combinations.

\subsubsection{Scenario A}

In this scenario, the first measurement would be done at $\Delta_1 =
\pi/2$ (with $L/E_\nu$ given by Eq.~(\ref{eq:sind})).  As
discussed earlier, this choice isolates the $\sin\delta$ term, removes
$\theta_{13}$ from the $(\delta,\theta_{13})$ ambiguity, and the
remaining $(\delta,\pi-\delta)$ ambiguity does not mix $CPC$ and $CPV$
solutions. These $L/E_\nu$ values also give a large $\nu_\mu$
disappearance, which facilitates the precision measurement of $\delta
m^2_{31}$ and $\sin^22\theta_{23}$. The baseline $L$ should be large
enough to avoid the sgn($\delta m^2_{31}$) ambiguity ($L \gsim 2000$~km
assures this for $\sin^2 2\theta_{13}\gsim 0.01$).
Representative values of $L$ and $E_\nu$ are given in
Table~\ref{tab:Delta}. Measuring $P(\nu_\mu \to \nu_e)$ and $\bar
P(\bar\nu_\mu \to \bar\nu_e)$ at one such $L$
and $E_\nu$ should determine sgn($\delta m^2_{31}$), $\theta_{13}$
(modulo the $\theta_{23}$ ambiguity, if present), and whether or not
$CP$ is violated (as discussed in Sec.~\ref{sec:theta23}, the existence
of a $\theta_{23}$ ambiguity will not give a large amount of $CPC/CPV$
confusion).

The second measurement should be one that best resolves the $(\delta,
\pi-\delta)$ ambiguity. In principle, $\Delta_2 = \pi$, which
eliminates the $\sin\delta$ terms in the probabilities and leaves only
$\cos\delta$ terms, gives the maximal separation of $\delta$ and $\pi -
\delta$. Thus, the first measurement gives $\sin\delta$, the second
gives $\cos\delta$, from which the value of $\delta$ may be
inferred. Furthermore, if $\theta_{13}$ is determined from the first
measurement, then both $P$ and $\bar P$ would not have to be measured in
the second measurement; one is sufficient to determine $\delta$. Whether
one used neutrinos or antineutrinos in the second measurement would be
determined by which gave the larger event rate, taking into account
neutrino fluxes, cross sections, and oscillation probabiltities. If
$\delta m^2_{31} > 0$, then neutrinos would be best for the second
measurement, due to the larger flux and cross section; for $\delta
m^2_{31} < 0$, antineutrinos may be the better choice if the matter
enhancement is enough to overcome the lower flux and cross section for
antineutrinos. If both the first and second measurements are done at the
same $L$, then $\Delta_2 = \pi$ means that the appropriate energy in the
second experiment is $E_2 = E_1/2$.

In practice, there are other values $\Delta_2$ that are not close to
$\pi/2$ that could potentially work for the second
measurement. The optimal $\Delta_2$ also depends on the particular
values of $f$, $\bar f$ and $g$ at the various $L$ and $E_\nu$, as well
as on neutrino parameters that are currently unknown ($\delta
m^2_{21}$, $\theta_{12}$, and $\theta_{13}$). We do not pursue the
optimization here.

\subsubsection{Scenario B}

\begin{figure}[t!]
\centering\leavevmode
\psfig{file=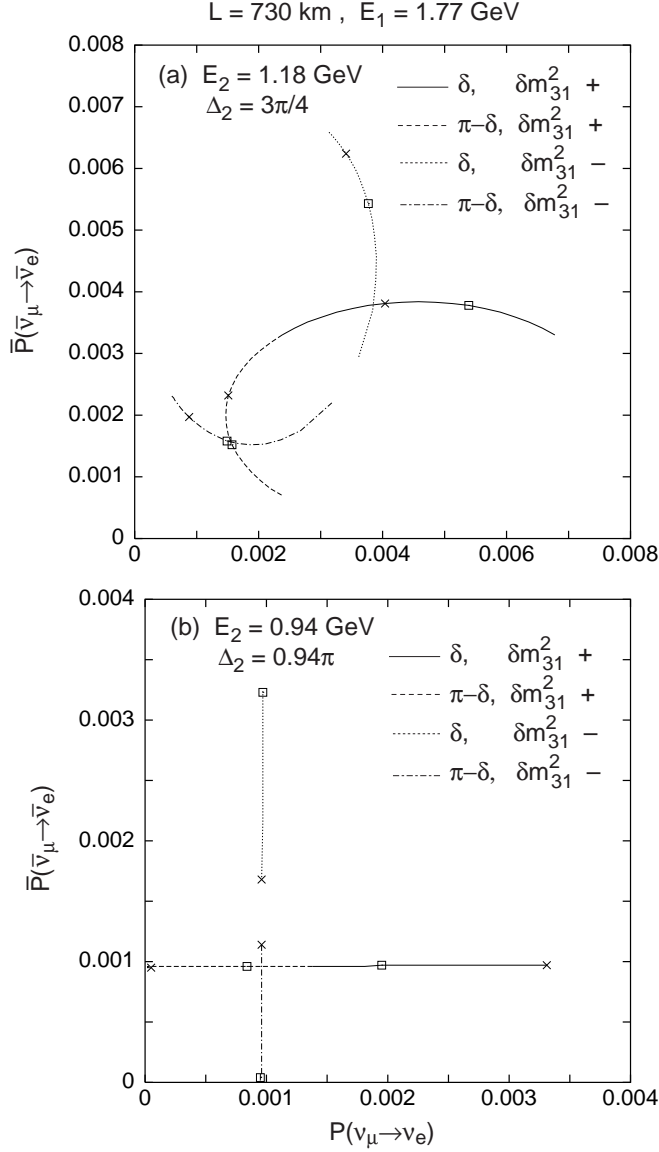,width=3.5in}
\medskip
\caption{Values of $P$ and $\bar P$ in a second measurement when there
is a four--fold degeneracy in the first measurement for (a) $E_2 =
1.18$~GeV ($\Delta_2 = 3\pi/4$) and (b) $E_2 = 0.94$~GeV ($\Delta_2 =
\pi/(1+\hat A) = 0.94\pi$), with $L_1 = L_2 = 730$~km and $E_1 =
1.77$~GeV ($\Delta_1 = \pi/2$). Each curve (solid, dashed, dotted,
dash--dotted) represents one of the four
solutions that are degenerate. Points labelled by the same symbol
(crosses or boxes) correspond to solutions that are degenerate with each
other in the measurement at $L_1$ and $E_1$.}
\label{fig:break4}
\end{figure}

If the first measurement is done at an $L$ that is not large enough to
resolve the sgn($\delta m^2_{31}$) ambiguity, then the second
measurement must be tailored to both determine sgn($\delta m^2_{31}$)
and resolve the $(\delta,\pi-\delta)$ ambiguity, i.e., it must break a
four--fold degeneracy. As discussed above, the choice $\Delta_2 = \pi$
determines $\delta$, but it can be shown that at shorter $L$ an
approximate degeneracy with parameters of the opposite sgn($\delta
m^2_{31}$) remains (e.g., see Fig.~\ref{fig:sign}c). Another example is
shown in Fig.~\ref{fig:break4}a, where the near degeneracy of parameters
with the opposite sgn($\delta m^2_{31}$) remains for some values of
$\delta$ (while the crosses in Fig.~\ref{fig:break4}a are
well--separated in the second measurement, the boxes are not). However,
at $\Delta_2 = \pi/(1\pm\hat A)$, either $\bar f$ or $f$ vanishes
(depending on the sign of $\delta m^2_{31}$), and the four ambiguous
solutions occupy four separate regions in $(P,\bar P)$ space, as shown
in Fig.~\ref{fig:break4}b.  Although the four regions in
Fig.~\ref{fig:break4}b overlap somewhat, when the point of one
degenerate solution is in the overlap region the points of the other
three degenerate solutions are not (the crosses and boxes are always
well--separated). Thus the four--fold ambiguity involving $(\delta,
\pi-\delta)$ and sgn($\delta m^2_{31}$) will always be resolved. A
disadvantage of Scenario~B is that because either $f$ or $\bar f$ is
zero in the second measurement, $P$ and $\bar P$ tend to be smaller, so
that event rates may be somewhat lower than for other values of
$\Delta$.

% 8

Some examples are given in Table~\ref{tab:break4}. For instance, if
the first baseline is $L_1=730$~km (Fermilab to Soudan), then $\Delta_1
= \pi/2$ for $E = 1.77$~GeV, and two possibilities for a second
experiment with the same beam energy are $L_2=1295$ and $1700$~km, which
serendipitously are very close to the distances from Fermilab
to Homestake and from Fermilab to Carlsbad (or Brookhaven to Soudan).

\begin{table}[t]
\squeezetable
\caption[]{Possible sets of neutrino beam energies and baselines that
will resolve the four--fold parameter ambiguity when the measurements
are done at shorter $L$ (such that the sgn($\delta m^2_{31}$) ambiguity
is not resolved in either experiment).}
\label{tab:break4}
\begin{tabular}{c|c|c}
Fixed $L$ (km) & $E_1$ (GeV) & $E_2$ (GeV) \\
\hline
300 & 0.73 & 0.355, 0.375\\
730 & 1.77 & 0.835, 0.940\\
\hline
Fixed $E$ (GeV) & $L_1$ (km) & $L_2$ (km) \\
\hline
0.73 & 300 & 575, 630\\
1.77 & 730 & 1295, 1700\\
\end{tabular}
\end{table}

In practice, narrow band beams are not monoenergetic.  However, values
of $\Delta_2$ close to $\pi/(1\pm\hat A)$ also give reasonably good
separation of the ambiguities, as long as $\Delta_2$ is not close to
$\pi$. If the fractional beam spread is more than $|\hat A|$, a slightly
different average value of $\Delta_2$ might be preferable, to ensure
that no significant part of the beam has $\Delta_2$ too close to $\pi$.

\begin{figure}
\centering\leavevmode
\psfig{file=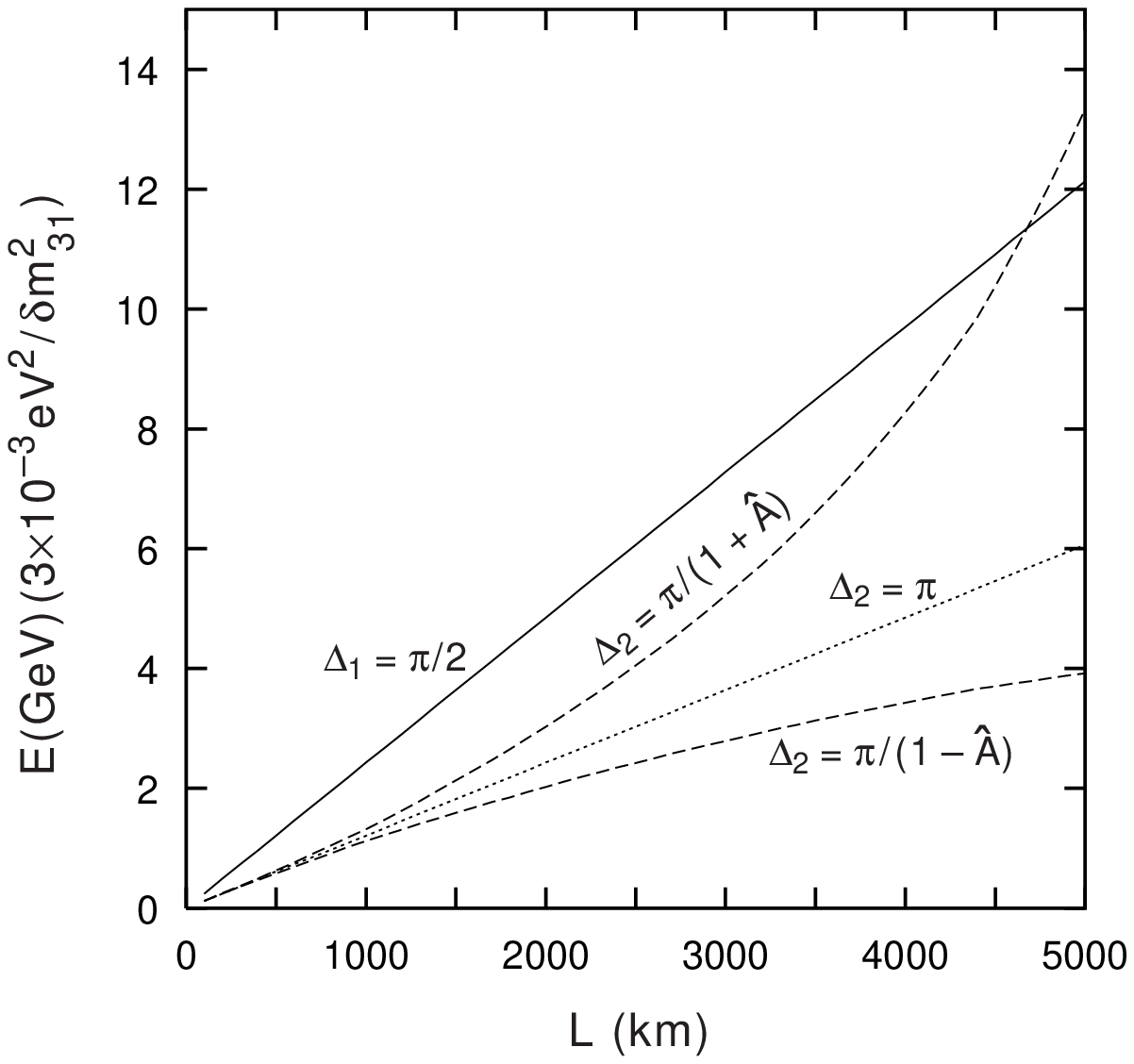,width=3.5in}
\medskip

\caption{Values of $L$ and $E_\nu$ for a first measurment at
$\Delta_1 = \pi/2$ (solid curve) and a second measurement at $\Delta_2
= \pi$ (Scenario A, dotted) or $\pi/(1\pm\hat A)$ (Scenario B, dashed),
which breaks the parameter degeneracy in each case.}
\label{fig:EvsL}
\end{figure}

Fig.~\ref{fig:EvsL} summarizes the possibilities for Scenerios A and B,
showing $E_\nu$ versus $L$ for the first measurement
done at $\Delta_1 = \pi/2$ (solid curve) and a possible second
measurement at $\Delta_2 = \pi$ (Scenario A, dotted curve) or
$\pi/(1\pm\hat A)$ (Scenario B, dashed curves).

% 9

\subsubsection{Scenario C}

This scenario uses the fact that the probabilities are insensitive to
the parameters of the $\delta m^2_{21}$ scale at $L \simeq 7600$~km, as
noted in Sec.~\ref{sec:orbits}. If the first measurement of $P$ and
$\bar P$ were done at $L \simeq 7600$~km, $\theta_{13}$ would be
determined (modulo the $\theta_{23}$ ambiguity), and because the
distance is large enough sgn($\delta m^2_{31}$) would also be determined
from the large matter effect. A second measurement of $P$ and $\bar P$
could then be done at an $L$ and $E_\nu$ such that $\Delta =
(2n-1)\pi/4$, which gives the maximum ``fatness'' of the orbit
ellipse~\cite{minakata} (see Sec.~\ref{sec:orbits}), which in turn
should best distinguish different values of $\delta$.
One disadvantage of Scenarios~B and C compared to Scenario~A is that
both $P$ and $\bar P$ must be determined in both measurements.

%In Scenario~C it may be possible that only $P$ (or $\bar P$, depending
%on sgn($\delta m^2_{31}$)), would have to be measured at $L \simeq
%7600$~km, while $P$ and $\bar P$ would be measured at $\Delta = (2n-1)
%\pi/4$. In this case the measurement at $\Delta = (2n - 1)\pi/4$ would
%have to be done first, and at a large enough $L$ that the matter
%effects allowed a determination of sgn($\delta m^2_{31}$).

\subsubsection{Discussion of scenarios}

Although the three scenarios discussed above are not necessarily the only
solutions to the ambiguities, in each case one measurement is chosen to
eliminate one or more of the parameters from the ambiguities, leaving
the second measurement to resolve only the remaining ambiguities. In
this sense, they are cleaner measurements. Scenario~A would appear to be
more favorable, since in principle the first measurement alone could
determine $\sin\delta$, sgn($\delta m^2_{31}$), and $\theta_{13}$
(modulo the $\theta_{23}$ ambiguity), and thus determine whether or
not $CP$ is violated.  Also, as discussed in Sec.~\ref{sec:theta23},
even if there is a $\theta_{23}$ ambiguity, the magnitude of the
$CPC/CPV$ confusion appears to be relatively small for the usual range
of neutrino parameters considered.

\subsubsection{Implication for JHF experiments}

\begin{figure}
\centering\leavevmode
\psfig{file=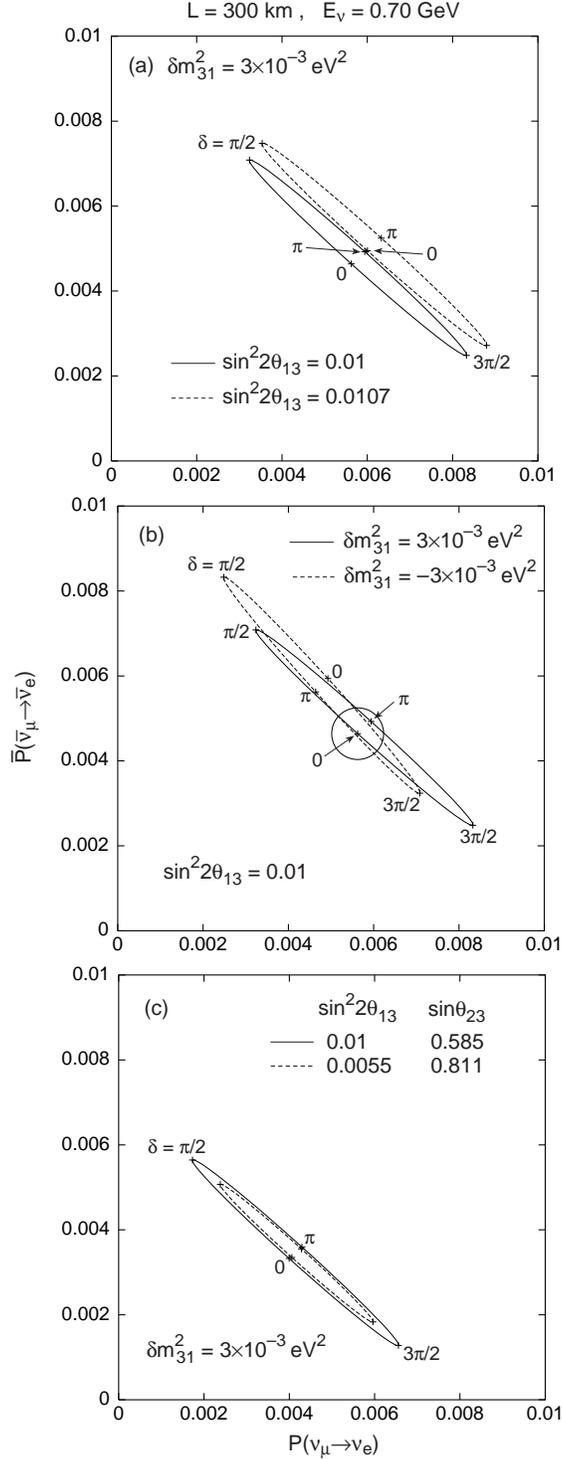,width=3.0in}
\medskip

\caption[]{Examples of the three types of ambiguities for the proposed
SuperJHF--HyperK experiment~\cite{jhfsk} with $L = 300$~km and $E_\nu =
0.7$~GeV:
(a) ($\delta, \theta_{13}$) ambiguity,
(b) sgn($\delta m^2_{31}$) ambiguity, and
(c) ($\theta_{23}, \pi/2 - \theta_{23}$) ambiguity.
In each case $\delta m^2_{21} = 5\times10^{-5}$~eV$^2$,
$\sin^22\theta_{23} = 1$, and $\sin^22\theta_{12} = 0.8$, unless
otherwise stated in the figure. The circle in (b) indicates the size of
the expected experimental uncertainties~\cite{jhfsk}.}
\label{fig:jhfsk}
\end{figure}

The proposed SuperJHF--HyperKamiokande experiment~\cite{jhfsk} satisfies the
requirements for the first experiment of Scenario B. The plan is to have
a neutrino energy such that $\Delta$ is at the first peak of the
oscillation for $L=300$~km; if $\Delta$ is not exactly on the peak
(e.g., if $\delta m^2_{31} = 3\times10^{-3}$~eV$^2$), a long narrow
ellipse results instead of a straight line (see Fig.~\ref{fig:jhfsk}).
Because the distance is relatively short, the sgn($\delta m^2_{31}$)
ambiguity is not likely to be resolved since there is considerable 
overlap of the two sgn($\delta m^2_{31}$) ellipses (see
Fig.~\ref{fig:jhfsk}b). For example, for $\sin^22\theta_{13} = 0.01$ the
point for $\delta = 0$ with $\delta m^2_{31} > 0$ is nearly the same as
the point for $\delta = 1.18\pi$ for $\delta m^2_{31} < 0$
(Eq.~\ref{eq:sindprime}, which measures the size of the $CPC/CPV$
confusion for the sgn($\delta m^2_{31}$) ambiguity, gives the about the
same numerical result). The expected 90\%~C.L. uncertainty in $\delta$
in this case is about $0.07\pi$ near $\delta = 0$~\cite{jhfsk}, so we
see that the sgn($\delta m^2_{31}$) ambiguity caused by the matter
effect would seriously impede a proper measurement of 
$\delta$, although there is the possibility that the
SuperJHF--HyperK experiment might measure a point $(P,\bar P)$ that was
sufficiently outside the overlap region, thereby determining sgn($\delta
m^2_{31}$)~\cite{minakata}. A possible ($\theta_{23}, \pi/2 -
\theta_{23}$) ambiguity also remains, which could lead to a
corresponding ambiguity in $\theta_{13}$, as shown in
Fig.~\ref{fig:jhfsk}c.

Even though JHF may not sit exactly on the peak of the oscillation
(i.e., $\Delta = \pi/2$), Fig.~\ref{fig:jhfsk} shows that the three
ambiguities discussed here are present. Also, Fig.~\ref{fig:jhfsk}a
shows that the ambiguity in $\theta_{13}$ is relatively small (of order
10\% or less) if $\Delta$ is close, but not exactly equal, to $\pi/2$.
Thus as long as $L/E_\nu$ is chosen so that the oscillation is close to
the first peak, we expect that the scenarios discussed here for
determining the neutrino mass and mixing parameters will be valid.

% 9.5

\section{Summary}
\label{sec:summary}

There is an eight-fold, $(\delta\,, \theta_{13})$-sgn$(\delta
m^2_{31})$-$(\theta_{23}\,, \pi/2-\theta_{23})$ degeneracy affecting the
neutrino mixing matrix determined in long-baseline neutrino oscillation
experiments. If $\sin^2 2\theta_{23}$ is almost unity as is favored by
current Super-K and K2K data, this is reduced to a four-fold ambiguity.
To break this four-fold ambiguity to a simple $(\delta, \pi-\delta)$
ambiguity which does not interfere with a determination of whether or
not $CP$ is violated in the neutrino sector, we find that an experiment
should be performed at the first oscillation maximum corresponding to
$\Delta=\pi/2$ and at a baseline of at least about 1300--2000~km,
depending on the value of $\delta m^2_{21}$.
Representative values of $L$ and $E_\nu$ that yield $\Delta=\pi/2$ are
given in Table~\ref{tab:Delta}.

The obvious advantages of choosing $\Delta=\pi/2$ are that
$\nu_\mu\rightarrow\nu_e$ transitions are nearly maximal even when
matter effects are accounted for and the $\nu_\mu\rightarrow \nu_\tau$
oscillation (which has small matter effects) is maximal, allowing a
precise measurement of $\sin^2 2\theta_{23}$ and $\delta m^2_{31}$.  By
choosing $\Delta=\pi/2$, the $(\delta\,,\theta_{13})$ degeneracy
represented by the $P$-$\bar P$ ellipse collapses to a line leaving a
$(\delta, \pi-\delta)$ ambiguity (see Fig.~\ref{fig:delta}) which
unambiguously determines whether or not $CP$ is violated.
%It is extremely important to note that this occurs independently of the
%$(\theta_{23}\,, \pi/2-\theta_{23})$ ambiguity because the $x^\prime=x$
%ambiguity (which contains the $(\theta_{23}\,, \pi/2-\theta_{23})$ ambiguity)
% is decoupled from the $(\delta, \pi-\delta)$ ambiguity.
The central reason for the choice of the first oscillation peak over
other peaks is that practically speaking, the sgn$(\delta m^2_{31})$
ambiguity can be resolved only for this peak (see Fig.~\ref{fig:minx}).
The other peaks succeed in eliminating this ambiguity only for the
smallest values of $\delta m^2_{21}$ in the LMA region and for $\sin^2
2\theta_{13}$ close to the CHOOZ bound. As shown in
Figs.~\ref{fig:deltasign1}-\ref{fig:deltasign2} to remove this ambiguity
simultaneously with the $(\delta\,, \theta_{13})$ ambiguity requires
that the baseline be at least 1300 km for $\sin^2 2\theta_{13}>0.01$ and
$\delta m^2_{21} = 5\times10^{-5}$~eV$^2$. For lower values of $\sin^2
2\theta_{13}$ and/or higher values of $\delta m^2_{21}$, longer
baselines than this 
are needed.  The exciting aspect of an experiment at $\Delta=\pi/2$ and
a sufficiently long baseline is that all degeneracies other than the
$(\theta_{23}\,, \pi/2-\theta_{23})$ degeneracy can be broken to a
harmless $(\delta, \pi-\delta)$ ambiguity with only a single baseline
and energy.  The remaining $(\delta, \pi-\delta)$ ambiguity can be
removed by making a second measurement at $\Delta=\pi$ which leaves only
$\cos \delta$ terms in the probabilities and provides the maximal
separation between $\delta$ and $\pi-\delta$.  The $(\theta_{23}\,,
\pi/2-\theta_{23})$ degeneracy cannot be eliminated even with
measurements at a second baseline and energy because in the leading term
in $P(\nu_\mu \to \nu_e)$ and ${\bar P}(\bar\nu_\mu \to \bar\nu_e)$,
$\sin2\theta_{13}$ is paired with $\sin\theta_{23}$ (see
Eqs.~(\ref{eq:P}) and (\ref{eq:Pbar})).  Fortunately, the mixing of the
$CPC$ and $CPV$ solutions arising from this degeneracy are of order or smaller
than the experimental uncertainty in $\delta$, thereby making it less severe.
Only a neutrino factory, which offers the unique ability to compare
$P(\nu_e \to \nu_\mu)$ and $P(\nu_e \to \nu_\tau)$, can disentangle
$\sin2\theta_{13}$ from $\sin\theta_{23}$ and find whether $\theta_{23}$
is less than or greater than $\pi/4$.

If it is not possible to have an experiment with $L$ sufficiently large
to find sgn$(\delta m^2_{31})$, a second experiment is necessary to
simultaneously resolve the $(\delta, \pi-\delta)$ ambiguity and determine
sgn$(\delta m^2_{31})$. One possibility is to chose $\Delta=\pi$, but as 
suggested by Fig.~\ref{fig:sign}c, a sgn$(\delta m^2_{31})$
ambiguity may still remain when experimental errors are included. 
If $\Delta = \pi/(1\pm \hat A)$, either $f$ or $\bar f$
vanishes, depending on sgn$(\delta m^2_{31})$, and the four-fold
degeneracy breaks into four separate regions as in Fig.~\ref{fig:break4}b.
See Table~\ref{tab:break4} for some examples of how this scenario can be
implemented. The proposed SuperJHF--HyperK experiment would satisfy the
requirements for the first measurement of this type; it has the
limitation of a possible sgn($\delta m^2_{31}$) confusion that leads to
an ambiguity in the value of $\delta$, which may compromise its ability
to unambiguously establish $CP$ violation.

In Fig.~\ref{fig:EvsL} we summarize the baselines and energies for two
measurements, one at $\Delta_1 = \pi/2$ and another at either $\Delta_2
= \pi$ (if the first measurement can determine sgn$(\delta m^2_{31})$
and only the $(\delta, \pi-\delta)$ ambiguity needs resolution) or
$\Delta_2 = \pi/(1\pm \hat A)$ (if the first measurement can not be
performed at a long enough baseline and the four-fold degeneracy
$(\delta, \pi-\delta)$-sgn$(\delta m^2_{31})$ needs to be broken).
Another possibility is to have one measurement at $L \simeq 7600$~km
with $\hat A \Delta_1 = \pi$ and a second measurement with $\Delta_2
\simeq (2n-1)\pi/4$. If K2K, MINOS, ICARUS and OPERA find that
$\theta_{23}$ is not very close to $\pi/4$, a neutrino factory will be
needed to resolve the $(\theta_{23}\,, \pi/2-\theta_{23})$ ambiguity.

In our analysis we have assumed that $\delta m^2_{31}$ and
$\sin^22\theta_{23}$ are known when the experiments described here are
done. In fact, they will likely be determined only to 10\% or so. Once
these uncertainties are included the minimum value of $L$ required to
resolve the sgn($\delta m^2_{31}$) ambiguity, e.g., in Scenario~A, could
be slightly longer than indicated here. Also, because $\delta m^2_{31}$
is not precisely known, the average neutrino energy will not necessarily
be exactly at the peak defined by $\Delta = \pi/2$. However, as our
analysis of the proposed SuperJHF--HyperK experiment shows, only minimal
uncertainties in $\delta$ and $\sin^22\theta_{13}$ are introduced by
these factors, and the three principal ambiguities discussed in this
paper will be qualitatively unchanged.

\section*{Acknowledgments}

This research was supported in part by the U.S.~Department of Energy
under Grants No.~DE-FG02-95ER40896, No.~DE-FG02-01ER41155 and 
No.~DE-FG02-91ER40676, and in
part by the University of Wisconsin Research Committee with funds
granted by the Wisconsin Alumni Research Foundation.

\appendix
\section{}

We provide a complete set of analytic expressions for the
off-diagonal probabilities that are valid
in the regime $|\hat A|> |\alpha|$, which roughly
translates to $E_\nu >0.5$ GeV.
The diagonal probabilities can easily be found from them. The off-diagonal
probabilities are for a normal hierarchy 
(in addition to Eqs.~(\ref{eq:P}) and~(\ref{eq:Pbar}))
\begin{eqnarray}
P(\nu_e \to \nu_\tau) = {\rm cot}^2 \theta_{23} x^2 f^2 - 2 x y f g
(\cos\delta\cos\Delta + \sin\delta\sin\Delta)
+ {\rm tan}^2 \theta_{23} y^2 g^2\,,
\label{eq:Petau}\\
{\bar P}(\bar{\nu}_e \to \bar{\nu}_\tau) ={\rm cot}^2 \theta_{23} x^2 \bar f^2
- 2 x y \bar f g (\cos\delta\cos\Delta
- \sin\delta\sin\Delta) + {\rm tan}^2 \theta_{23} y^2 g^2 \,,
\label{eq:Petaubar}
\end{eqnarray}
and
\begin{eqnarray}
P(\nu_\mu \to \nu_\tau) &=& \sin^2 2\theta_{23} \sin^2\Delta
\nonumber \\ && + \alpha
 \sin 2\theta_{23} \sin 2\Delta \bigg({\hat A \over 1-\hat A}
\sin \theta_{13} \sin 2\theta_{12} \cos 2\theta_{23} \sin\Delta-\Delta
\cos^2 \theta_{12} \sin 2\theta_{23}\bigg)\,.
\label{eq:Pmutau}
\end{eqnarray}
For ${\bar P}(\bar{\nu}_\mu \to \bar{\nu}_\tau)$, replace $\hat A$ by
$- \hat A$ in Eq.~(\ref{eq:Pmutau}). Note that $P(\nu_\mu \to \nu_\tau)$
is independent of $\delta$ to ${\cal O}(\alpha)$. To obtain the probabilites
for an inverted hierarchy, the transformations
$\hat A \to - \hat A$,
$\alpha \to -\alpha$ and $\Delta \to -\Delta$ must be made 
(implying $f \leftrightarrow -\bar f$ and $g \to -g$ in 
Eqs.~(\ref{eq:Petau}-\ref{eq:Petaubar}), and for the $T$-reversed channels
the sign of the $\sin\delta$ term must be changed. 

We now compare the results of the analytic expressions with the numerical
integration of the evolution equations of neutrinos through the Earth.
We integrate the equations along a neutrino path using a Runge-Kutta method.
The step size at each point along the path is 0.1\% of the shortest
oscillation wavelength given by the scales $\delta m_{31}^2$ and $A$.
We account for the dependence of the density on depth by using the
Preliminary Reference Earth Model (PREM)~\cite{PREM}.
To calculate the analytic probability, we use the average value of
the electron density along the neutrino path. We provide some values 
in Table~\ref{neavg} for the reader's use; they are not indicative of the
precision with which the electron density is known. We include
subleading $\theta_{13}$ effects, which however are not relevant for
$\sin^22\theta_{13}$ of ${\cal O}(0.01)$ or smaller.
They are of importance at $\theta_{13}$ for which the CHOOZ limit
$\sin^22\theta_{13}<0.1$ (at 95\% C. L.) is saturated.
The parameters chosen to make this comparison are $\delta m_{31}^2=3.5 \times
10^{-3}$ eV$^2$, $\delta m_{21}^2=5 \times 10^{-5}$ eV$^2$,
$\theta_{23}=\pi/4$, $\theta_{12}=\pi/6$,
$\sin^22\theta_{13}=0.01$ and $E_\nu=5$ GeV.
Thus, the ensuing comparison
is not affected by dropping subleading $\theta_{13}$ effects.
A normal mass hierarchy ($\delta m^2_{31} > 0$)
is assumed. We will comment on the comparison involving an inverted mass
hierarchy.

\begin{table}[t]
\begin{eqnarray}
\begin{array}{lccr}
\rm{Baseline}\, (\rm{km}) &  & \langle N_e \rangle  \\
\hline
   0-800    &&        1.284400 \\    
   900      &&        1.320264 \\   
   1000     &&        1.356722 \\  
   1100     &&        1.374506 \\   
   1200     &&        1.474359 \\   
   1300     &&        1.516450 \\   
   1400     &&        1.543904 \\   
   1500     &&        1.563676 \\   
   1600     &&        1.578441 \\   
   1700     &&        1.590218 \\   
   1800     &&        1.599484 \\   
   1900     &&        1.607012 \\   
   2000     &&        1.613276 \\   
   2100     &&        1.618752 \\   
   2200     &&        1.622966 \\   
   2300     &&        1.626689 \\   
   2400     &&        1.630101 \\   
   2500     &&        1.633021 \\   
   2600     &&        1.635155 \\   
   2700     &&        1.636799 \\   
   2800     &&        1.638902 \\   
   2900     &&        1.640515 \\   
   3000     &&        1.641341 \\   
   3100     &&        1.642923 \\   
   3200     &&        1.643719 \\   
   3300     &&        1.644499 \\   
   3400     &&        1.653481 \\   
   3500     &&        1.659367 \\   
   3600     &&        1.663974 \\   
   3700     &&        1.668762 \\   
   3800     &&        1.672956 \\   
   3900     &&        1.676851 \\   
   4000     &&        1.681093 \nonumber  
\end{array}
\end{eqnarray}
\caption[]{The average values of the electron density $\langle N_e \rangle$
 at baselines for which the analytic approximations of the probabilities 
accurately represent the numerical integration of the evolution equations.}
\label{neavg}
\end{table}

\begin{figure}[h]
\centering\leavevmode
\psfig{file=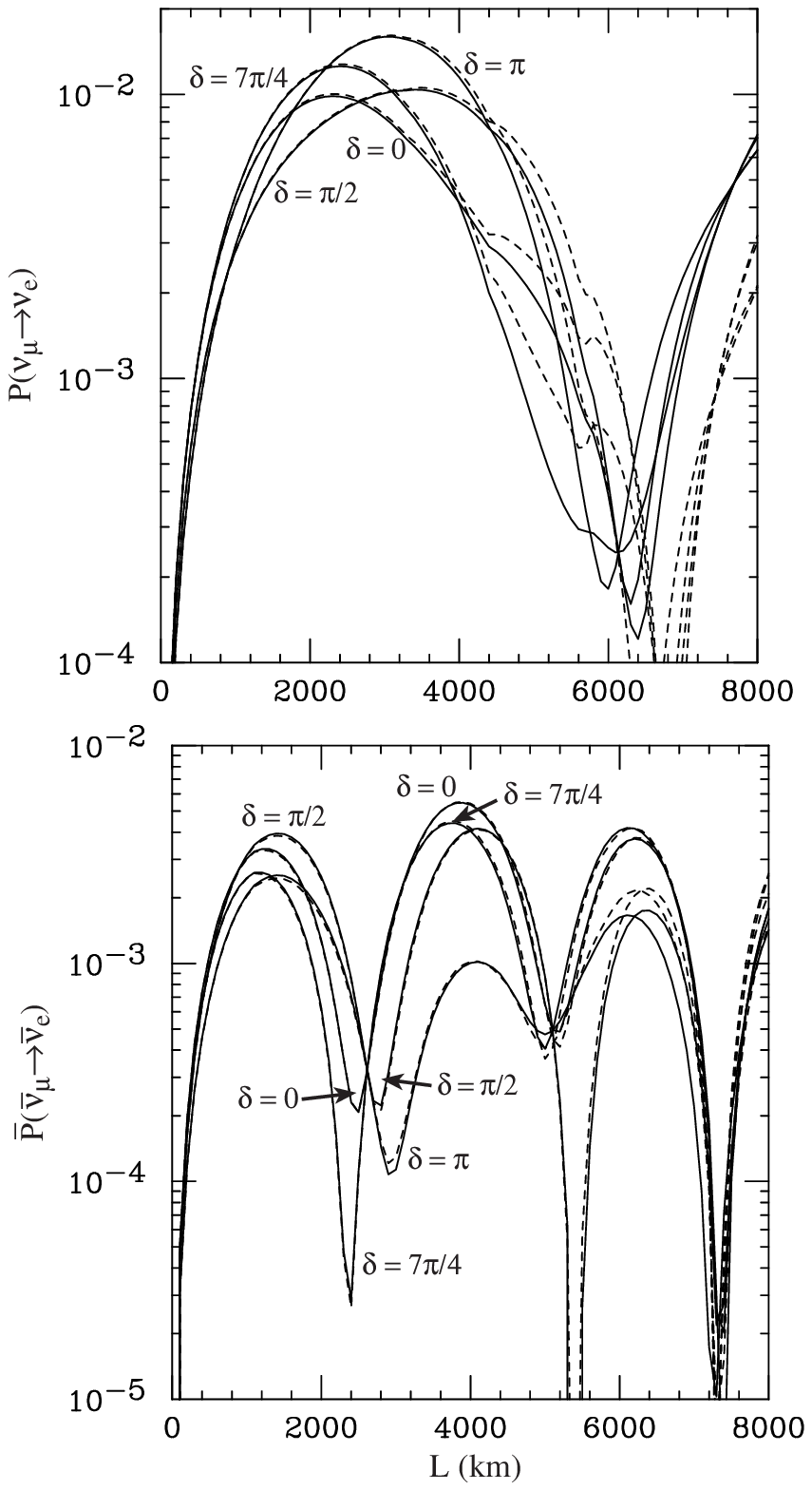,width=3.5in}
\medskip

\caption{$P(\nu_\mu \to \nu_e)$ and 
${\bar P}(\bar{\nu}_\mu \to \bar{\nu}_e)$ versus $L$
for $\delta=0,\pi/2,\pi,7\pi/4$.
The agreement between the analytic
formulae (solid lines) and the numerical results (dashed lines)
is excellent for
distances up to about 4000 km. The parameters chosen to make this
comparison are $\delta m_{31}^2=3.5 \times
10^{-3} {\rm eV^2}$, $\delta m_{21}^2=5 \times 10^{-5} {\rm eV^2}$,
$\theta_{23}=\pi/4$, $\theta_{12}=\pi/6$,
$\sin^22\theta_{13}=0.01$ and $E_\nu=5$ GeV.}
\label{fig:comparison1}
\end{figure}

Figure~\ref{fig:comparison1} shows $P(\nu_\mu \to \nu_e)$ and
$P(\bar{\nu}_\mu \to \bar{\nu}_e)$ versus distance for $\delta=0,
\pi/2, \pi$, and $7\pi/4$. The agreement between the
analytic formulae (solid lines) and the numerical results (dashed lines)
is excellent for distances up to about 4000 km. Beyond that, the overlap
between the lines degrades and for $L \gsim 5000$ km, the analytic
equation completely breaks down. The analytic expression for
$P(\bar{\nu}_\mu \to \bar{\nu}_e)$ works for much longer distances than
that for $P(\nu_\mu \to \nu_e)$. Analogously, for the inverted mass
hierarchy, $P(\nu_\mu \to \nu_e)$ is valid to longer distances than
$P(\bar{\nu}_\mu \to \bar{\nu}_e)$.

% 10

Reference~\cite{freund} claims good agreement between the analytic and
numerical results for $L$ even larger than 10000 km when a constant
density is assumed for the Earth's density profile. The use of a
realistic density profile as in the PREM model shows that the agreement
deteriorates at much smaller distances.

\begin{figure}[h]
\centering\leavevmode
\psfig{file=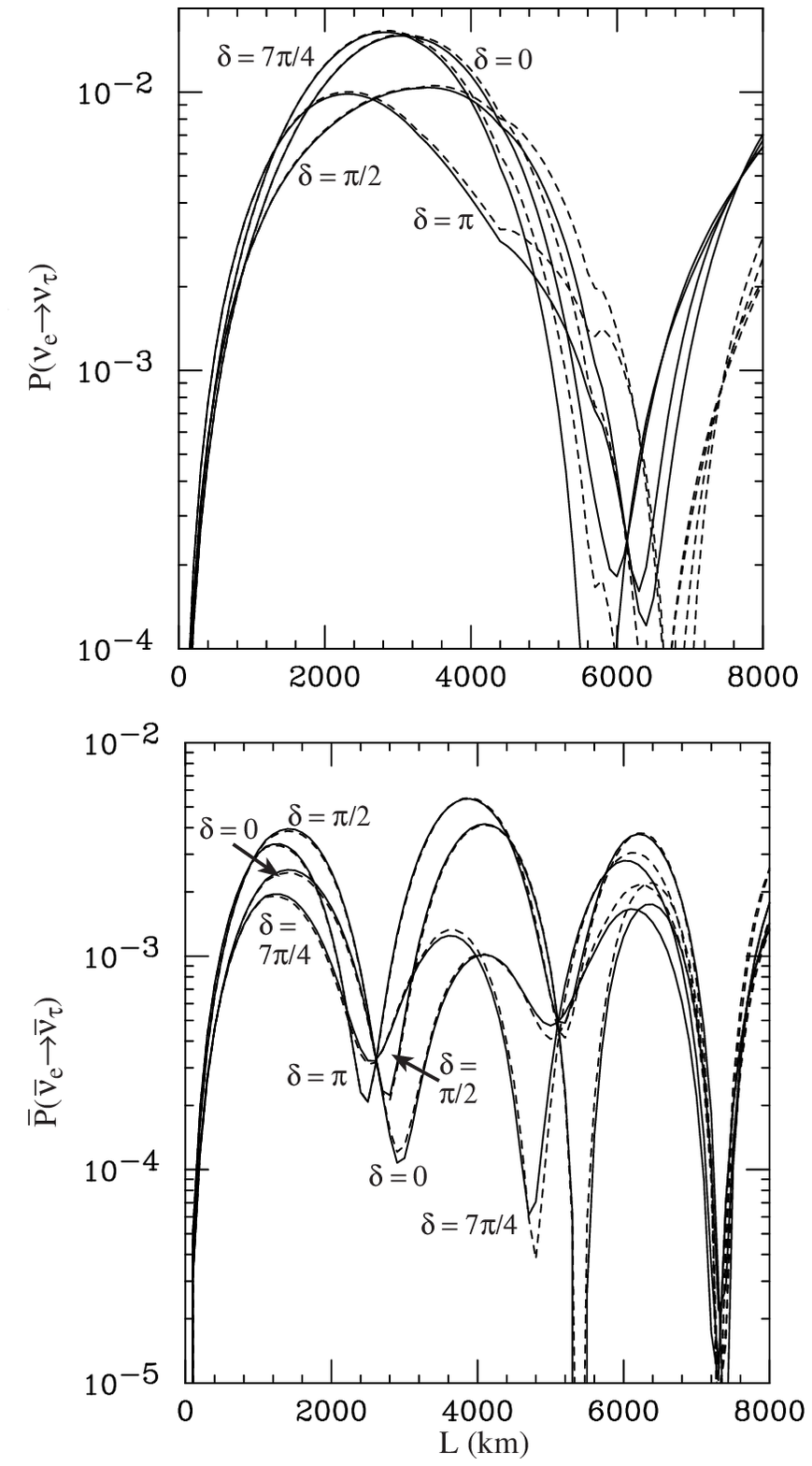,width=3.5in}
\medskip

\caption{$P(\nu_e \to \nu_\tau)$  and 
${\bar P}(\bar{\nu}_e \to \bar{\nu}_\tau)$ versus $L$
for the same set of parameters as in Fig.~\ref{fig:comparison1}.}
\label{fig:comparison2}
\end{figure}

\begin{figure}
\centering\leavevmode
\psfig{file=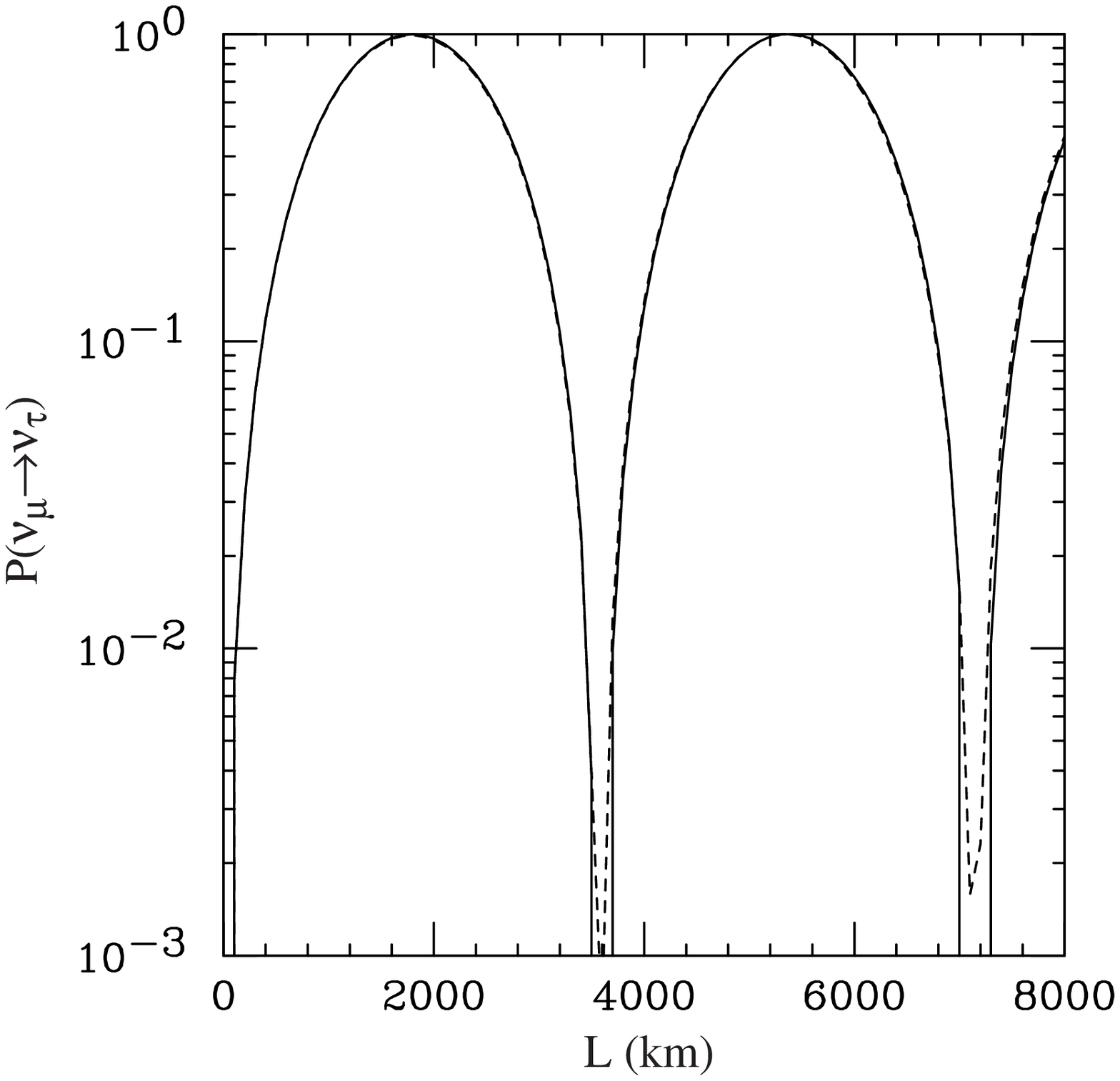,width=3.5in}
\medskip

\caption{$P(\nu_\mu \to \nu_\tau)$ versus $L$. The
analytic expression (solid line) and the numerical calculation (dashed line)
agree almost exactly for the entire range in $L$.
${\bar P}(\bar{\nu}_\mu \to \bar{\nu}_\tau)$ is almost identical to
$P(\nu_\mu \to \nu_\tau)$ because of insignificant matter effects.}
\label{fig:comparison3}
\end{figure}

For the sake of completeness we display the corresponding comparisons for
$P(\nu_e \to \nu_\tau)$, ${\bar P}(\bar{\nu}_e \to \bar{\nu}_\tau)$ and
$P(\nu_\mu \to \nu_\tau)$ in Figs.~\ref{fig:comparison2} and~\ref{fig:comparison3}.
The parameter values chosen are the same as for
Fig.~\ref{fig:comparison1}. The range of validity of
Eqs.~(\ref{eq:Petau}) and~(\ref{eq:Petaubar}) is the same as for
Eqs.~(\ref{eq:P}) and~(\ref{eq:Pbar}). However, Eq.~(\ref{eq:Pmutau}) agrees
almost exactly with the numerical result for the entire range considered.
This is because matter effects are very small in comparison to the
leading contribution. For the same reason,
${\bar P}(\bar{\nu}_\mu \to \bar{\nu}_\tau)$ is almost identical to
$P(\nu_\mu \to \nu_\tau)$.

% 11

% 12

\begin{figure}
\centering\leavevmode
\psfig{file=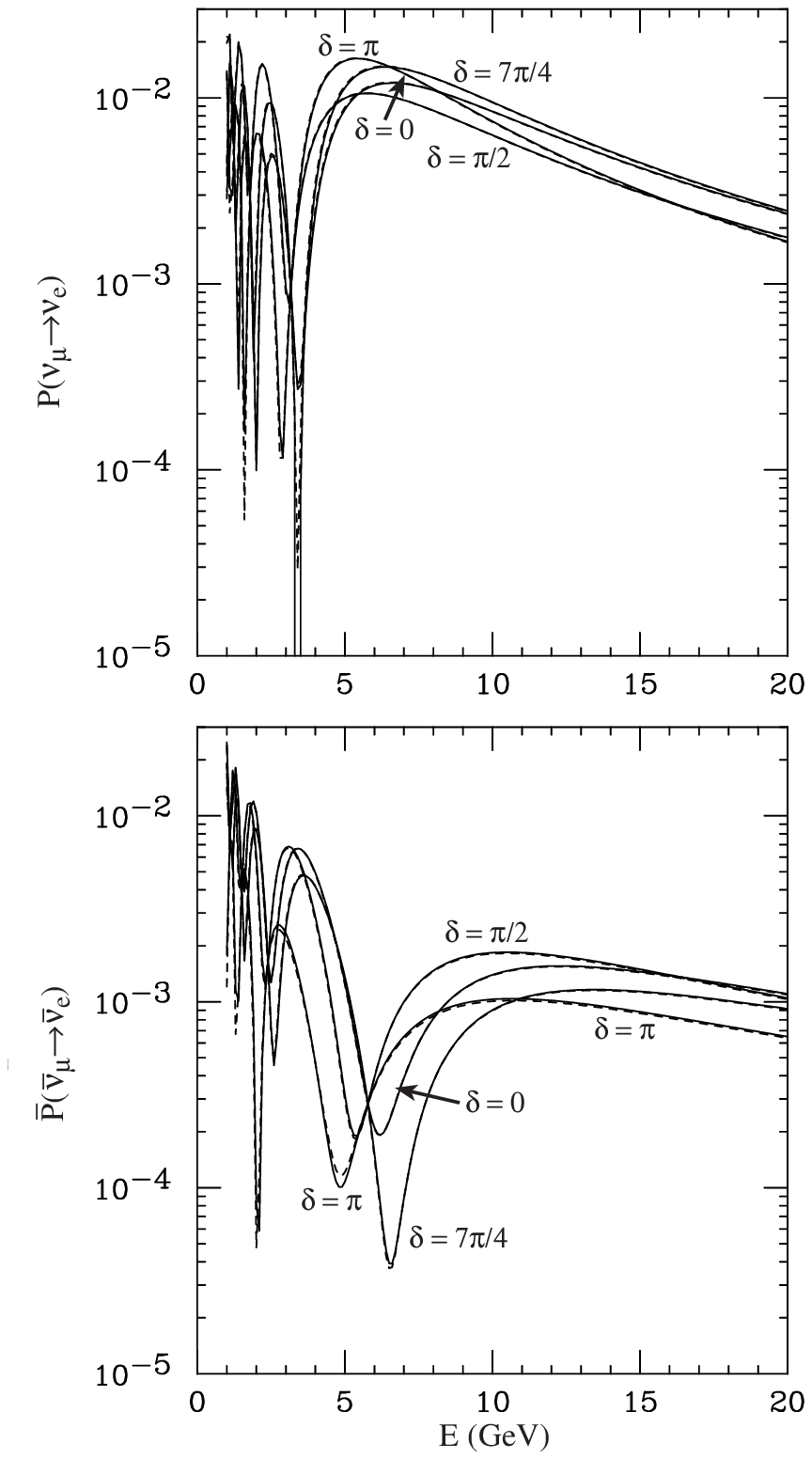,width=3.5in}
\medskip

\caption{$P(\nu_\mu \to \nu_e)$ and
${\bar P}(\bar{\nu}_\mu \to \bar{\nu}_e)$ vs $E_\nu$ for $L = 2900$~km. The
oscillation parameters are the same as in Fig.~\ref{fig:comparison1}.
The solid lines (analytic equations) and dashed lines (numerical evaluation)
are almost undistinguishable.}
\label{fig:comparison4}
\end{figure}

In Fig.~\ref{fig:comparison4}, we show how well the analytic
probabilities $P(\nu_\mu \to \nu_e)$ and ${\bar P}(\bar{\nu}_\mu \to
\bar{\nu}_e)$ agree with the numerical integration for $L=2900$~km (the
longest baseline emphasized in this work) as a function of neutrino
energy. The oscillation parameters used are the same as for
Fig.~\ref{fig:comparison1}. The precision is remarkable
for the spectrum of energies of interest. For shorter baselines, the
agreement gets even better.

% 13

\begin{figure}
\centering\leavevmode
\psfig{file=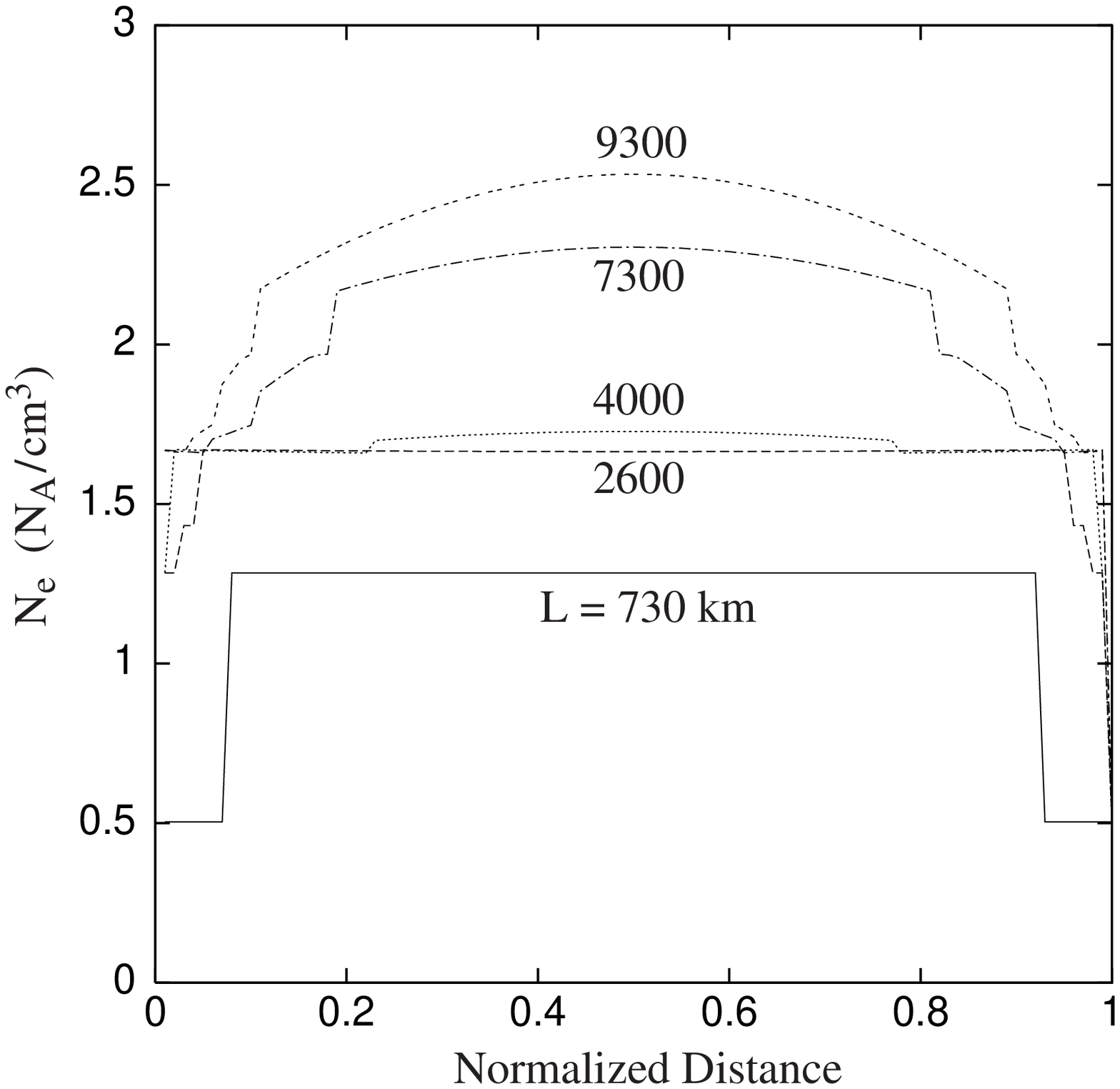,width=3.5in}
\medskip

\caption{Density profiles along a selection of of chords of length $L$ passing
through the Earth; the horizontal axis is the fraction of the total path length.
}
\label{fig:14}
\end{figure}

We now make some cautionary remarks. Our comparisons were made for
$\alpha=0.0143$, the parameter in which the series was expanded, and
$\sin^22\theta_{13}=0.01$ which is assumed to be no greater than of
${\cal O}(\alpha)$. These values are motivated by the existing reactor
bounds and global fits to the atmospheric and solar data. However, as
either of these parameters gets larger, the agreement between the
analytic equations and the numerical results deteriorates at long
baselines even if subleading $\theta_{13}$ effects are included.
Conversely, the agreement improves with smaller values of $\alpha$ and
$\theta_{13}$.  As a rule of thumb, we recommend that the constant
density approximation to the probabilities be used only for distances
less than 4000 km. As can be seen from Fig.~\ref{fig:14}, for $L<4000$
km, the density profile is nearly constant for most of the neutrino
path, thereby satisfying the implicit assumption (of a constant density
profile) under which analytic probabilities are valid.

% 12

We have stated that the analytic expressions are accurate for $E_\nu >
0.5$ GeV for baselines of 4000-5000 km. This robust bound can be relaxed
for $L \lsim 350$ km to $E_\nu$ as low as 0.05 GeV. However, for such low
values of $E_\nu$, the sensitivity of the analytic probabilities to
$\alpha$ and $\theta_{13}$ is high and care must be taken in their use.
For example, a comparison with a numerical integration is desirable if
$\alpha$ and $\theta_{13}$ are relatively large.

\end{document}